\documentclass{article}
 \usepackage{authblk}
\usepackage{graphicx} 
\usepackage{amsmath, amsfonts, amssymb, amsthm}
\usepackage[letterpaper,top=2cm,bottom=2cm,left=2cm,right=2cm,marginparwidth=1.75cm]{geometry}
\usepackage{subcaption}

\usepackage{float}
\usepackage[dvipsnames]{xcolor}
\usepackage{cleveref}

\newcommand{\Ro}{\mathrm{Ro}}
\newcommand{\Fr}{\mathrm{Fr}}

\newcommand{\td}{\text{d}}

\newcommand{\Tr}{\mathrm{Tr}}

\title{Minimum-enstrophy solutions in topographic quasi-geostrophic flow on the rotating sphere}
\author[1]{Sagy Ephrati\thanks{Corresponding author: \texttt{s.ephrati@imperial.ac.uk}}}
\author[2]{Erik Jansson \thanks{E-mail address: \texttt{eoj23@cam.ac.uk}}}

\affil[1]{Department of Mathematics, Imperial College London}
\affil[2]{Department of Applied Mathematics and Theoretical Physics, University of Cambridge}

\date{\today}

\begin{document}

\maketitle

\begin{abstract}

    The minimum-enstrophy theory of Bretherton and Haidvogel postulates that two-dimensional turbulent systems evolve to a state that minimises enstrophy at a fixed energy level. 
    We extend this to the rotating spherical quasi-geostrophic setting, accounting for bottom topography and the fully nonlinear Coriolis effect, resulting in latitude-dependent effects not present in planar approximations. 
    We prove existence and nonlinear stability of minimum-enstrophy solutions and describe analytically asymptotic regimes for certain rates of rotation, topography scales, and energy values.
    We compute the minimum-enstrophy solutions by a structure-preserving method for the quasi-geostrophic equations on the sphere.
    We apply the method to a range of parameter values, including those describing Jupiter's atmosphere.
    The results reveal a distinct latitude dependence of the flow, with a tendency for topographical trapping near the poles and zonal flow near the equator, depending on the chosen parameters.
    The predicted nonlinear stability is confirmed numerically by integrating perturbed solutions using a structure-preserving time discretisation.
\end{abstract}

\section{Introduction}
Large-scale planetary flows, including atmospheric and oceanic dynamics, are driven by an interplay between rotation, spherical geometry, and topography.
In two-dimensional turbulence, the dynamics are governed by the nonlinear advection of potential vorticity (PV), leading to a double cascade and a self-organisation of the flow into coherent large-scale structures.
According to the selective decay hypothesis of Bretherton and Haidvogel \cite{bretherton1976two}, such systems evolve to a state that minimises enstrophy while nearly preserving energy.
This concept has been extensively studied using planar approximations, but these simplified geometries neglect latitude-dependent effects that characterise flows on rotating spheres.
The aim of this work is to study the qualitative effects of rotation and spherical geometry on minimum-enstrophy solutions and their dependence on topography.
In doing so, we extend the theory of minimum-enstrophy solutions from the torus to the spherical setting while accounting for the fully nonlinear Coriolis effect, and highlight similarities and differences in the resulting large-scale flow features.

The advection of PV gives rise to the double cascade observed in two-dimensional turbulence.
In the absence of viscous dissipation, external forcing, and damping, many models possess conserved quantities of which the kinetic energy and the enstrophy are often deemed most important. 
Nonlinear advection causes different scales of motion to interact with each other via triadic interactions.
To conserve energy and enstrophy, it is necessary that energy removed from the median mode moves to the lower wavenumber mode, while enstrophy from the median mode moves to the higher wavenumber mode \cite{zeitlin2018geophysical}.
The inverse energy cascade leads to self-organisation of the flow, and might ultimately lead to large-scale condensate formation \cite{modin2020casimir, boffetta2012two}.
The downward enstrophy cascade induces turbulent flow at successively higher wavenumbers, until the flow fluctuations ultimately reach a scale that is sufficiently small for viscous dissipation to take place.
This dissipation decreases the total enstrophy yet leaves the total energy nearly unchanged, a distinction which is referred to as \textit{selective decay} \cite{bretherton1976two}.
For this reason, Bretherton and Haidvogel argued that enstrophy must dissipate as long as the direct enstrophy cascade is maintained, and they therefore studied \textit{minimum-enstrophy solutions} that minimise the enstrophy for a given value of the energy. 
This framework provides insight into the PV dynamics, in particular with respect to topographic features of the flow domain.

Minimum-enstrophy solutions have been used for studying the interaction between topography and PV over the past decades. 
These solutions are equilibria to the quasi-geostrophic (QG) equations, and exhibit distinct qualitative behaviour where the PV aligns with topography, as already noted by \cite{bretherton1976two}. 
Additional early results on equilibrium states of quasi-geostrophic models were derived using statistical mechanics \cite{salmon1976equilibrium, carnevale1987nonlinear}, focusing on $f$-plane and $\beta$-plane approximations of the sphere.  
The analysis of minimum-enstrophy solutions was extended to the shallow water equations by \cite{sanson2010evolution}, who demonstrated that the flow aligns with the topography profile in decaying turbulence.
The role of topography has been further formalised by \cite{siegelman2023two}, who observed that the extent to which the PV is affected by topography depends on the kinetic energy of the solution.
At higher energies, the topography is a weak perturbation to the merger of vortices often observed in flat-bottom two-dimensional turbulence, whereas PV follows the topography profile at lower energies.
A similar conclusion was reached in the analysis of persistent vortices over bathymetry \cite{lacasce2024vortices}.
Recent work numerically investigated  the effect of topography roughness on decaying turbulence, and found that topographically trapped vortices appeared more frequently on rougher bottom profiles \cite{priya2026two}.
The applicability of inviscid minimum-enstrophy solutions to weakly forced and dissipated flows was studied by \cite{gallet2024two}, who derived a branch enstrophy-minimizing solutions at sufficiently large energy values and showed that selective decay quantitatively predicts condensate sizes.
We contribute to this body of literature by providing a formal derivation and numerical characterization of these solutions in a spherical setting, accounting for the interplay between curvature, rotation and topography.

Due to the spherical geometry, global quasi-geostrophic turbulence exhibits a pronounced latitudinal dependency.
This is reflected in the Rossby deformation radius, which increases toward the equator, and the Rhines length, which decreases.
These variations result in a latitude-dependent scale for the arrest of the inverse energy cascade \cite{rhines1975waves, salmon1982geostrophic}.
Furthermore, they establish a critical latitude below which eddies can elongate to form zonal structures and above which turbulence becomes increasingly isotropic \cite{theiss2004equatorward, franken2024critical}.
Similarly, we show that minimum-enstrophy solutions also display a distinct dependence on latitude: flow is topographically trapped near the poles, whereas these effects diminish near the equator to allow the emergence of zonal flow.

The principle of selective decay has also been observed in magnetohydrodynamics (MHD), where conserved quantities are present akin to those in geophysical fluid dynamics.
For example, in two-dimensional MHD, conservation of the mean-square magnetic potential can be observed while the energy dissipates \cite{biskamp2001two}.
In both geophysical and plasma contexts, this process can be understood as a relaxation toward an equilibrium state that is constrained by the dynamic invariants of the system.
This is exploited in the derivation of metric brackets in MHD, which induce dynamics where the Hamiltonian (total energy) remains invariant while the entropy grows, thereby providing thermodynamically consistent dissipation that aims to maximize entropy \cite{morrison1986paradigm}.
The correspondence between maximal-entropy and minimum-enstrophy solutions is made explicit in \cite{brands1999maximum}, which assesses the applicability of the latter to statistical theory.
Notably, a linear relationship between the vorticity and the stream function is found to be a reliable approximation to the equilibrium state derived from statistical theory.
Here, we utilize this variational framework to study the spatial structure of a class of equilibria in quasi-geostrophic flow on the sphere.

\begin{figure}[ht]
    \centering
    \includegraphics[width=0.7\linewidth]{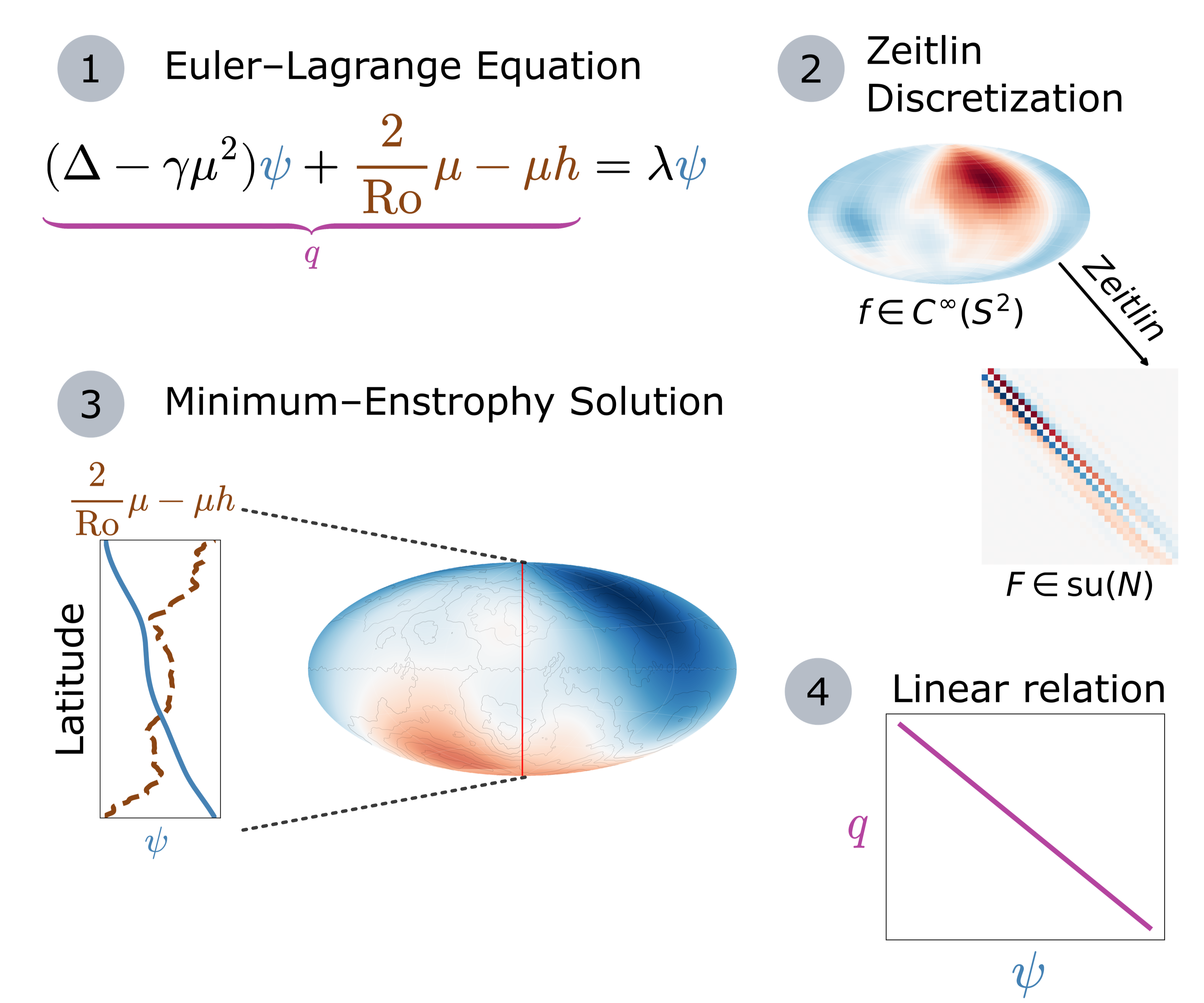}
    \caption{Graphical summary of the paper. A necessary condition for enstrophy minimisation yields an Euler--Lagrange equation. This equation is solved numerically using Zeitlin's method.
    The resulting minimum-enstrophy solution may be aligned with the topography or with the rotation, depending on the parameter values. The Euler--Lagrange equations imply a linear relationship between stream function and potential vorticity; this is confirmed numerically.}
    \label{fig:ga}
\end{figure}

The primary objective of this work is to study minimum-enstrophy solutions for spherical flows
by extending the theory of Bretherton and Haidvogel to spherical geometry.
Specifically, we derive and analyse minimum-enstrophy solutions for QG on a rotating sphere and characterise these solutions in asymptotic regimes based on rotation, topography scales, and energy.
We prove existence and nonlinear stability of the minimum-enstrophy solution.
We describe an algorithm to numerically compute these solutions to compare qualitative flow features for a range of parameters and assess the stability of the found solutions.
A graphical summary of the paper is provided in Figure \ref{fig:ga}.

The paper is structured as follows.
The spherical QG equations are introduced in Section \ref{sec:governing_equations}.
The variational problem for finding minimum-enstrophy solutions is laid out in Section \ref{sec:minens}, followed by proofs of existence and stability of the solutions in Section \ref{subsec:existence_stability} and a study of asymptotic regimes in Section \ref{subsec:asymptotic_regimes}.
In Section \ref{sec:computational_method}, we highlight Zeitlin's structure-preserving numerical method, here adapted to computing minimum-enstrophy solutions, which we subsequently employ in a sequence of numerical tests in Section \ref{sec:numerical_demonstrations}.
The paper is concluded in Section \ref{sec:conclusions}.

\section{The spherical QG equations}\label{sec:governing_equations}
In this paper, we study the spherical quasi-geostrophic equations (QGE) on the sphere.
These equations arise as a balanced approximation of the single-layer shallow water equations (SWE) on the sphere through several simplifying steps, which we summarise below.
For full details, we refer to the works of Verkley \cite{verkley2009balanced} and Schubert et al. \cite{schubert2009shallow}, who independently derived this model, using previous results by Kuo \cite{kuo1959finite} and Charney and Stern \cite{charney1962stability}.
For a model derivation including the Lagrangian and Hamiltonian formulation of the spherical QGE, we refer to \cite{luesink2024geometric}.

The shallow water equations (SWE) describe fluid motion in shallow domains, such as the oceans or the atmosphere. 
The equations are derived from the three-dimensional Navier--Stokes equations via several simplifying assumptions, including a skewed ratio between characteristic horizontal and vertical length scales and velocities, negligible hydrostatic pressure effects, and incompressibility.
After depth-averaging the equations of motion, the resulting model consists of two velocity or momentum components and a free surface height.
The SWE furthermore permit a variable bottom topography (also referred to as bathymetry in oceanic context and orography in atmospheric context) and the extension to multi-layer models.

The QGE are derived from the SWE via several additional steps.
Starting from the single-layer SWE, taking the (two-dimensional) curl and the horizontal divergence of the momentum equations yields evolution equations for the potential vorticity (PV) and divergence.
By assuming a sufficiently small horizontal divergence, the horizontal velocity field can be approximated by the curl of a stream function. 
This approximation is based on Daley's simplest geostrophic relationship \cite{daley1983linear}, and also follows by assuming that fluctuations in the free surface height and the topography are small.
The assumption of small divergence subsequently allows us to simplify the divergence evolution to a balance equation which, by linearising, yields the linear balance equation. 
The small fluctuations in the height of the fluid column additionally permit expressing the PV in terms of the rotation, free surface height, and topography.
Finally, the linear balance equations are used in conjunction with the geopotential, expressed in terms of the free surface height, to define the relation between the PV and the stream function.
As such, the QGE provide a workable yet effective model for large-scale atmospheric and oceanic flows where rotation effects are important, by approximating the three-dimensional Navier--Stokes equations near geostrophic balance.

The resulting equations in dimensionless form read
\begin{align}
q_t &+ \{\psi, q\} = 0, \label{eq:qg1} \\
q &= \left(\Delta - \gamma\mu^2\right)\psi + \frac{2\mu}{\Ro} - \mu h, \label{eq:qg2}
\end{align}
where $q$ is the PV, $\psi\colon \mathbb{S}^2 \to \mathbb{R}$ is the stream function, $h$ is the topography, $\Delta$ is the Laplace--Beltrami operator on the sphere, and $\mu=\cos \theta$, where $\varphi \in [0, \pi]$ and $\theta \in[0, 2\pi]$ denote the colatitude and the longitude, respectively.
We assume $\varphi = 0$ to be the north pole and $\varphi=\pi$ the south pole.

Equation \eqref{eq:qg1} describes the advection of potential vorticity, here written in terms of the Poisson bracket $\{\cdot, \cdot\}$.
This is a bilinear skew-symmetric operator defined as \begin{equation}
    \{\psi, q\} = \nabla^\perp\psi\cdot\nabla q, \label{eq:poisson_bracket}
\end{equation}
where $\nabla^\perp$ is the gradient rotated by 90 degrees on the surface of the sphere.
For brevity, in the remainder of the paper we denote the inhomogeneous Helmholtz operator by \begin{equation}
    H\psi:= (\Delta -\gamma\mu^2)\psi.
\end{equation}
The dimensionless parameters are the Rossby number \begin{equation}
    \Ro =\frac{U}{\Omega L}
\end{equation}  and Lamb's parameter \begin{equation}
    \gamma = \frac{4\Omega^2 L^2}{gH_f},
\end{equation}
where $U$ denotes the characteristic velocity, $\Omega$ is the rotation frequency of the sphere, $L$ is the characteristic length scale, $g$ denotes the gravitational acceleration, and $H_f$ is the average fluid depth.
The Rossby number is the ratio between the characteristic velocity and the rotation velocity, where small values indicate increased importance of rotation effects in the dynamical system. 
Lamb's parameter is alternatively expressed as $\gamma = 4 (\mathrm{Fr/\Ro})^2$. Here, the Froude number $\mathrm{Fr}= U / \sqrt{gH}$ is the ratio between the characteristic velocity and the velocity of the fastest gravity wave.
Small values of $\gamma$ indicate that the dynamics are governed predominantly by stratification effects, whereas large values indicate dynamics dominated by rotation effects.
Estimates of the characteristic values and corresponding dimensionless numbers are given in  \Cref{tab:physical_parameters}, see also \cite{luesink2024geometric}.

\begin{table}[h]
    \centering
    \caption{Table of characteristic values and corresponding dimensionless numbers for Earth's ocean, Earth's atmosphere, and Jupiter's atmosphere. The table is adapted from \cite{luesink2024geometric}.}
    \begin{tabular}{l|c|c|c}
         & Atmospheric (Earth) & Oceanic (Earth) & Atmospheric (Jupiter) \\
         \hline
         Length $L$ [m]                         & $4.0\times 10^7$    & $3\times10^6$       &  $4.4\times10^8$\\
         Fluid depth $H_f$ [m]                  & $2.5\times 10^3$    & $10^2$              & $1.5\times10^5$  \\
         Velocity $U$ [ms$^{-1}$]               & $10^{-1}$           & $10^{-1}$           & $10^2$ \\
         Rotation frequency $\Omega$ [s$^{-1}$] & $1.2\times 10^{-5}$ & $1.2\times 10^{-5}$ & $2.8\times10^{-5}$ \\
         Gravitational acceleration $g$ [ms$^{-2}$] & $9.8$           & $9.8$               & $2.5\times 10^1$ \\
         \hline
         Rossby number $\Ro$                    & $2.1\times 10^{-4}$ & $2.8\times10^{-3}$  & $8.9\times10^{-3}$ \\
         Froude number $\Fr$                    & $6.4 \times10^{-4}$ & $3.2\times 10^{-3}$ & $5.2\times 10^{-2}$\\
         Lamb's parameter $\gamma$              & $3.8\times10^{1}$   & $5.3$ & $1.3\times 10^2$ \\
    \end{tabular}
    
    \label{tab:physical_parameters}
\end{table}

The equations \eqref{eq:qg1}-\eqref{eq:qg2} are similar to the barotropic vorticity equations on the $\beta$-plane, but the crucial difference is the fully nonlinear Coriolis term that includes latitude-dependent effects of rotation into the model.
The inhomogeneous (latitude-dependent) Helmholtz operator appearing in \eqref{eq:qg2} includes the Cressman stretching term $\gamma\mu^2\psi$, which modifies the non-divergent Rossby--Haurwitz wave frequency to the quasi-geostrophic Rossby--Haurwitz wave frequency, as extensively studied \cite{schubert2009shallow}.
This provides an alternative to empirical wave propagation factors dating back to the 1950s \cite{cressman1958barotropic}.

The QGE \eqref{eq:qg1}-\eqref{eq:qg2} may also be associated with Lagrangian and Hamiltonian functionals, and the dynamics can equivalently be stated in terms of a Lie--Poisson bracket \cite{luesink2024geometric}.
As a result, in the absence of external forcing and dissipation, the QGE conserve the total kinetic energy and any integrated analytic function of the PV, known as Casimir invariants, with enstrophy being a well-known example.
The total energy and enstrophy are respectively given by \begin{align}
    E[\psi] &= -\frac{1}{2}\int\! \psi H \psi \,\td A \label{eq:energy},\\
    \mathcal{E}[\psi] &= \frac{1}{2}\int\!\left(H\psi+\frac{2\mu}{\Ro}-\mu h\right)^2\,\td A, = \frac{1}{2}\int\!q^2\,\td A,   \label{eq:enstrophy}
\end{align}
where $\td A = \sin\theta\,\td\theta\,\td\varphi$.
The definition of the enstrophy in terms of $\psi$ is included specifically as it will be used in the definition of minimum-enstrophy solutions.

\section{Minimum-enstrophy solutions} \label{sec:minens}
The hypothesis that two-dimensional turbulent flows evolve toward a state of minimum enstrophy while nearly preserving total energy was first formalised by Bretherton and Haidvogel \cite{bretherton1976two}.
Following their framework, we derive conditions for flow patterns on a rotating sphere that minimise the enstrophy for a fixed energy value.
Mathematically, we treat this as the constrained minimisation problem
\begin{equation}
    \label{eq:minimisation_problem}
    \begin{split}
    \min_{\psi \in C^\infty_0(\mathbb{S}^2)} &\mathcal{E}[\psi] \\
    \text{subject to}\quad  &E[\psi] = E_*.
    \end{split}
\end{equation}

We approach this problem by deriving the Euler--Lagrange equations for problem \eqref{eq:minimisation_problem}.
To that end, we derive expressions for the variations of the enstrophy and energy, respectively denoted by $\delta \mathcal{E}$ and $\delta E$, by applying a variation $\eta$ to the stream function.
Specifically, a necessary condition for enstrophy minimisation is obtained by requiring that $\delta \mathcal{E} + \lambda\delta E$ vanishes for all variations of $\psi$, where $\lambda$ is a Lagrange multiplier.
We find
\begin{equation}
\begin{split}
    \delta \mathcal{E} + \lambda\delta E &= \int\!\eta\,H\Big[H\psi + \frac{2\mu}{\Ro} - \mu h\Big]\,\td A - \lambda \int\eta\,H\psi\,\td A \\
    &=\int\!\Big(H\Big[\underbrace{H[\psi] + \frac{2\mu}{\Ro}-\mu h}_{q}-\lambda\psi\Big]\Big)\eta\,\td A.
\end{split}
\label{eq:variation}
\end{equation}
The integral \eqref{eq:variation} should vanish for any variation $\eta$, and therefore $H\big[H[q-\lambda\psi]\big]=0$ is a necessary condition for the minimum-enstrophy solution.
Since the inhomogeneous Helmholtz operator has a trivial null space \cite{verkley2009balanced}, this requirement reduces to $q=\lambda\psi$. 
Therefore, the necessary condition for the minimum-enstrophy solution is expressed solely in terms of the stream function as \begin{equation}
    H\psi + \frac{2\mu}{\Ro}-\mu h = \lambda \psi.
    \label{eq:necessary_requirement}
\end{equation}
In the remainder of this section, we establish existence and uniqueness of solutions to the above equation, and investigate asymptotic solution regimes.

\subsection{Existence and stability of minimum-enstrophy solutions}\label{subsec:existence_stability}
Equation \eqref{eq:necessary_requirement} provides a necessary, but not sufficient requirement for minimum-enstrophy solutions. 
That is, a solution $\psi$ that satisfies \eqref{eq:necessary_requirement} is not guaranteed to minimise enstrophy for a given energy value.
A priori, we cannot ensure existence of solutions to \Cref{eq:necessary_requirement}, nor that any found solution is in fact a stable minimum-enstrophy solution. 
To address these issues, we use the spectral properties of the operator $H$ analogous to the analysis of \cite{gallet2024two}.
By expanding the stream function in the eigenbasis of $H$, we can constructively demonstrate the desired existence and stability results.

\paragraph{Expansion in spheroidal harmonics} The operator $H$ is self-adjoint and negative-definite, and it therefore possesses a sequence of eigenpairs $(\omega_k, e_k)$ such that the eigenvalues are strictly negative and decreasing, $\omega_0 > \omega_1 > \ldots >-\infty$. 
The corresponding eigenfunctions $\{e_k\}_{k=0,\ldots,\infty}$ form an orthonormal basis of square-integrable functions on $\mathbb{S}^2$. 
These eigenfunctions are exactly the \emph{spheroidal} harmonics, which can be used, e.g., to study dispersion properties of Rossby--Haurwitz waves in the spherical QGE \cite{schubert2009shallow}.
While these harmonics generally lack a simple closed-form expression and must be computed numerically \cite{schubert2009shallow}, we treat them abstractly to derive existence and stability of minimum-enstrophy solutions.

It is straightforward to express solutions to Eq. \eqref{eq:necessary_requirement} in the orthonormal eigenbasis $\{e_k\}_{k=1}^\infty$ of $H$.
For clarity, we let $\xi = \frac{2\mu}{\Ro}-\mu h$ so that Eq. \eqref{eq:necessary_requirement} is rewritten as
\begin{align} 
(H-\lambda)\psi = -\xi. \label{eq:xi_equation}
\end{align}
If $\lambda > \omega_0$, then the operator $H-\lambda$ is invertible, and we can compute a unique $\psi$. 
Expanding both $\xi$ and $\psi$ in the spheroidal harmonic basis as $\xi = \sum_{k=0}^\infty \xi_k e_k$ and $\psi = \sum_{k=0}^\infty \psi_k e_k$, we insert these expressions into Eq. \eqref{eq:xi_equation} to find the formal solution \begin{equation}
    \psi(\lambda) = \sum_{k=0}^\infty \frac{\xi_k}{\lambda - \omega_k} e_k =: \sum_{k=0}^\infty \psi_k e_k. \label{eq:minens_in_spheroidal_basis}
\end{equation}
This expression is analogous to the minimum-enstrophy stream function on the torus expressed in Fourier coefficients \cite{bretherton1976two, gallet2024two}.

\paragraph{Existence and stability}
The spectral representation \eqref{eq:minens_in_spheroidal_basis} allows us to prove existence and stability in three steps: first, by showing that a unique $\lambda$ exists for any target energy; second, by establishing nonlinear stability of the resulting equilibrium; and third, by confirming that this equilibrium indeed minimises the enstrophy.

\subparagraph{Existence}
We proceed by showing that for any target energy $E_* >0$, Eq. \eqref{eq:necessary_requirement} has a unique solution $(\psi_*, \lambda_*)$ where $\lambda_* \in (\omega_0, \infty)$ and $E[\psi_*] = E_*$. 
To see this, we obtain the energy corresponding to the minimum-enstrophy solution as a function of $\lambda$ by inserting the expression for $\psi$ \eqref{eq:minens_in_spheroidal_basis} into the energy \eqref{eq:energy}, \begin{equation}
    E[ \psi(\lambda)] = -\frac{1}{2}\sum_{k=0}^\infty \frac{\omega_k \xi_k^2}{(\lambda-\omega_k)^2}.
\end{equation}
On the interval $\lambda \in( \omega_0, \infty)$, we observe that $E[\psi(\lambda)] \to 0$ as $\lambda \to \infty$ and $E[\psi(\lambda)]\to \infty$ as $\lambda \to \omega_0$. 
Moreover, the energy is a monotonically decreasing function of $\lambda$ in this range as
\begin{align*}
    \frac{\mathrm{d}E}{\mathrm{d}\lambda} = \sum_{k=0}^\infty \frac{\omega_k \xi_k^2}{(\lambda-\omega_k)^3}<0
\end{align*}
given that $\omega_k < 0$ and $\lambda-\omega_k>0$.
As a consequence, by the intermediate value theorem, there is a unique $\lambda_* \in (\omega_0, \infty)$ such that $E[\psi_*] = E_*$, where $\psi_*=\psi(\lambda_*)$. This holds for each $E_* > 0$.

\subparagraph{Nonlinear stability} 
We continue by demonstrating stability, which sets the stage for showing that $\psi_*$ actually minimises the enstrophy.

The first step in proving stability is recognising that a stationary point of the Lagrangian \eqref{eq:necessary_requirement} necessarily is a stationary solution of the quasi-geostrophic equation, since $q = \lambda \psi$ implies that $\{q,\psi\} = 0$. 
To assess stability, we employ the energy-Casimir method suited for equilibria in ideal fluids \cite{holm1985nonlinear}, following the approach presented by \cite{carnevale1987nonlinear}. 
The underlying idea is that nonlinear stability of the stationary solutions is investigated using energy and conserved quantities as a Lyapunov function \cite{arnold1966priori}. 

We consider a perturbation $\eta$ to the stream function.
The aim is to show that $F_C(\eta):= \mathcal{E}[\psi_*+\eta]-\mathcal{E}[\psi_*] + \lambda_*(E[\psi_* + \eta]-E_*)$, which has a critical point at $\psi_*$, defines a positive-definite quadratic form at $\psi_*$ and remains bounded under perturbations. 
We define this positive-definite quadratic form by computing the energy and the enstrophy in the perturbed state as \begin{equation}
    \begin{split}
        F_C(\eta) &= \mathcal{E}[\psi_* + \eta]-\mathcal{E}[\psi_*] + \lambda_*(E[\psi_* + \eta]-E_*)=\frac{1}{2}\int (H\eta)^2 - \lambda_*\eta H \eta\,\td A.
    \end{split}
    \label{eq:pos_def_quad_form}
\end{equation}
The derivation of the expression above is provided in Appendix \ref{app:nonlinear_stability_form}.
Because $F_C$ is a combination of conserved quantities, its value remains constant under the QG dynamics.
Therefore, the perturbation remains bounded in the norm induced by $F_C$.
It is readily shown that $F_C$ is positive definite by employing the spheroidal harmonic basis. We expand $\eta$ in this basis as $\eta = \sum_{k=0}^\infty \eta_k e_k$ to find
\begin{align*} 
F_C(\eta) = \frac{1}{2}\sum_{k=0}^\infty \eta_k^2 \omega_k^2 - \frac{1}{2}\sum_{k=0}^\infty \lambda_* \eta_k^2  \omega_k  = \frac{1}{2}\sum_{k=0}^\infty \eta_k^2 \omega_k^2 - \lambda_* \eta_k^2  \omega_k  = \frac{1}{2}\sum_{k=0}^\infty \eta_k^2 \omega_k (\omega_k-\lambda_*). 
\end{align*}
Since $\lambda_* > \omega_0$, every term is positive and therefore $F_C(\eta) >0$ for all nonzero $\eta$. 
Thus, we have found that $F_C$ is a positive definite quadratic form in which perturbations remain bounded in the norm induced by $F_C$.  
Therefore, $\psi_*$ is a stable equilibrium.

\subparagraph{Enstrophy minimisation}
With the quadratic form $F_C$ in place, it is straightforward to verify that $\psi_*$ defines a minimum-enstrophy solution. Namely, let $\psi$ be such that $E[\psi] = E_*$, meaning it contains the same energy as the proposed minimum-enstrophy solution. 
If we consider the perturbation $\eta = \psi-\psi_*$, it holds that $F_C(\psi-\psi_*) = \mathcal{E}[\psi] -\mathcal{E}[\psi_*] > 0$.
This means in particular that
\begin{align*}
    \mathcal{E}[\psi] > \mathcal{E}[\psi_*],
\end{align*}
showing that $\psi_*$ defines a minimum-enstrophy solution. 

The above proposition contains an unknown $\omega_0$. 
The study by \cite{schubert2009shallow} provides some numerical indications of the value of $\omega_0$.
While it is possible to derive bounds for $\omega_0$ using perturbative techniques, for the purposes of this study, a numerical estimation as detailed in \Cref{sec:computational_method} is sufficient. 

\subsection{Asymptotic regimes of minimum-enstrophy solutions}\label{subsec:asymptotic_regimes}

To gain further insight into minimum-enstrophy solutions, it is instructive to examine cases where the equations simplify.
To that end, we expand the solutions into a spherical harmonic basis and identify asymptotic regimes. 
Here, the spherical harmonics are used as this allows for a straightforward interpretation of minimum-enstrophy solutions in terms of physical length scales.

We denote by $Y_{l,m}$ the complex spherical harmonic function of degree $l$ and order $m$ (see also Appendix \ref{app:gaunt_coefficients}). 
We recall that $\mu=\cos\theta=\sqrt{\frac{4\pi}{3}}Y_{1,0}$, $\mu^2 = \frac{2}{3}\sqrt{\pi}Y_{0,0}+\frac{4}{3}\sqrt{\frac{\pi}{5}}Y_{2,0}$, and $Y_{0,0} = \frac{1}{2}\sqrt{\frac{1}{\pi}}$.
By expanding the Helmholtz operator and taking the inner product with $Y_{l,m}$, we obtain \begin{equation}
        \left\langle\Delta\psi-\gamma\mu^2\psi-\lambda\psi, Y_{l,m}\right\rangle = \left\langle\mu h - \frac{2\mu}{\Ro}, Y_{l,m}\right\rangle, \end{equation}
which is further expanded as
\begin{equation}
        \psi_{l, m}\left(l(l+1)+\frac{\gamma}{3}+\lambda\right) +\frac{4\gamma}{3}\sqrt{\frac{\pi}{5}}\langle Y_{2,0}\psi, Y_{l,m}\rangle = \frac{2}{\Ro}\sqrt{\frac{4\pi}{3}}\langle Y_{1,0}, Y_{l,m}\rangle - \sqrt{\frac{4\pi}{3}}\langle Y_{1,0}h, Y_{l,m}\rangle.
    \label{eq:linear_system_steady_state}
\end{equation}
It holds that  $\langle Y_{1,0}, Y_{l,m}\rangle = 1$ only when $(l,m)=(1,0)$ and equals zero in all other cases. The terms $\langle Y_{0,1}h, Y_{l,m}\rangle$ and $\langle Y_{0,2}\psi, Y_{l,m}\rangle$ lead to a coupling between different modes of the stream function and the topography.
Expanding $\psi$ and $h$ into spherical harmonics shows that these terms involve the inner product of three spherical harmonics, referred to as Gaunt coefficients. 
These are further detailed in Appendix \ref{app:gaunt_coefficients}.
Most importantly, $\psi_{l', m'}$ are coupled to coefficients of $\psi$ with $l$-values between $l'-2$ and $l'+2$, and to topography coefficients with $l$-values between $l'-1$ and $l'+1$. 
Because of this coupling, it is evident that a straightforward closed-form solution cannot be found. Moreover, there are an infinite number of coefficients $\psi_{l,m}$. 
Thus, we instead rely on numerical approximations to fully study the minimum-enstrophy solutions. This is discussed in detail in \Cref{sec:computational_method}.

Three limiting parameter regimes are identified in which the system \eqref{eq:linear_system_steady_state} can be simplified: 
\begin{itemize}
    \item \textit{Low-energy approximation}: A high value of Lagrange multiplier corresponds to low energy \cite{siegelman2023two}, therefore a sufficiently low energy corresponds to $\lambda\gg l(l+1), \gamma$ and we may approximate Eq. \eqref{eq:linear_system_steady_state} by 
    \begin{equation}\label{eq:low_en}
        \lambda\psi_{l,m}  \approx \left\langle \mu h - \frac{2\mu}{\Ro}, Y_{l,m}\right\rangle \implies \psi \approx \left(\mu h- \frac{2\mu}{\Ro}\right)\lambda^{-1}.
    \end{equation}
    This means that, at a fixed latitude, the stream function is proportional to the topography.
    This reduces to $\psi=(h-2/\Ro)\lambda^{-1}\sim h\lambda^{-1}$ at the poles, recovering the result of \cite{bretherton1976two} that the streamlines follow the contour lines of the topography.
    Near the equator, $\mu$ vanishes yet its gradient grows and rotation increasingly affects the streamlines.
    
    Inserting the approximation \eqref{eq:low_en} for $\psi$ into the definition of the PV \eqref{eq:qg2} yields \begin{equation}
        q=H\left[\left(\mu h - \frac{2\mu}{\Ro}\right)\lambda^{-1}\right] + \frac{2\mu}{\Ro}-\mu h = \left(\lambda^{-1}H - I\right)\left[\mu h - \frac{2\mu}{\Ro}\right]\approx \frac{2\mu}{\Ro} - \mu h,
    \end{equation}
    where the final approximation assumes that $\lambda\gg1$. This suggests that an alignment of the stream function with the topography induces an alignment of the PV with the topography.
    
    \item \textit{Small-scale approximation}: At sufficiently small scales, we have $l(l+1)\gg \gamma,\lambda$ and the approximation to Eq. \eqref{eq:linear_system_steady_state} reads \begin{equation}
        l(l+1)\psi_{l, m} \approx \left\langle \mu h - \frac{2\mu}{\Ro}, Y_{l,m}\right\rangle \implies \Delta\psi\approx \mu h - \frac{2\mu}{\Ro}. \label{eq:low_energy_approximation}
    \end{equation}
    Inserting this into the definition of the PV yields $q\approx-\gamma \mu^2 \psi$.
    Therefore, in this approximation the advection of the PV becomes independent of the topography. 
    This agrees with the results on the torus \cite{bretherton1976two}, where fluid particles are swept over small-scale topographic features.
    The length scales at which this approximation is valid decrease when rotation effects increase (growing $\gamma$) or when the energy decreases (growing $\lambda$).

    \item \textit{Large-Lamb limit}:
    At sufficiently large values of the Lamb parameter, we use $\gamma\gg l(l+1), \lambda$ and find the approximation
    \begin{equation}
        \frac{\gamma}{3}\psi_{l, m} +\frac{4\gamma}{3}\sqrt{\frac{\pi}{5}}\langle Y_{2,0}\psi, Y_{l,m}\rangle \approx \left\langle \mu h - \frac{2\mu}{\Ro}, Y_{l,m}\right\rangle \implies -\gamma\mu^2\psi\approx \mu h - \frac{2\mu}{\Ro},
        \label{eq:approx_rapid_rotation}
    \end{equation}
    which implies that $q\approx\Delta \psi$. A large value of $\gamma$ can be realised in two distinct physical regimes, the \textit{rapid rotation approximation} and the \textit{small-depth approximation}. 
    Rewriting the stream function approximation as \begin{equation}
        \psi\approx -\frac{1}{\gamma\mu}\left(h-\frac{2}{\Ro}\right) \label{eq:large_lamb_limit_1}
    \end{equation}
    shows that at a fixed latitude, $\psi$ is directly proportional to the topography.
    While the magnitude of $\psi$ remains small away from the equator due to $\gamma$ appearing in the denominator, it exhibits high sensitivity near the equator where $\mu$ approaches zero.
    
    In the rapid rotation approximation, increasing the rotation frequency causes an increase of $\gamma$, and is likely to be accompanied by small values of the Rossby number.
    In this case, the term involving $\Ro$ in \eqref{eq:approx_rapid_rotation} dominates the right-hand side. 
    This suggests that the rotational effects are stronger than the topography, forcing the streamlines to align with the rotation direction and facilitating the development of zonal flow.
    The assumption $\gamma\gg\lambda$ is consistent with this limit, as the appearance of $\lambda$ in \eqref{eq:energy} suggests that the energy increases with $\gamma$.

    In the small-depth approximation, $\gamma$ attains a large value when the fluid depth $H_f$ decreases while the Rossby number remains constant.
    This causes the gravity wave speed $\sqrt{gH_f}$ to diminish and the Froude number to increase. 
    Intuitively, the fluid's ability to balance perturbations via vertical adjustments is reduced, making the flow increasingly sensitive to the topography.
    
\end{itemize}

A crucial difference between the results presented above and previous results on the torus is the latitude-dependence of the stream function.
The inclusion of the full spherical Coriolis term reveals a gradual transition from the polar regime, where topographic effects are prominent, to the equatorial regime, where inertial and zonal effects emerge regardless of topographic features.
This suggests that minimum-enstrophy solutions on a rotating sphere are inherently anisotropic in the latitudinal direction, a characteristic that is absent in traditional $\beta$-plane approximations.
In the general, non-asymptotic regime, we must resort to numerical calculations, which we describe below.

\section{Computational method}\label{sec:computational_method}

As highlighted in  \Cref{subsec:asymptotic_regimes}, numerical approximations are crucial for understanding minimum-enstrophy solutions and their relation with topography. 
Here, we employ a recently developed geometric discretization for the QGE on the sphere \cite{franken2024zeitlin} to numerically approximate these solutions.
The approach is a self-consistent spectral method referred to as \textit{Zeitlin's method}. 

Zeitlin's method is a specific geometric discretization for two-dimensional Euler and QG dynamics. 
Originally derived for Euler dynamics on the torus \cite{zeitlin1991finite},
it was subsequently extended to ideal flows on the (rotating) sphere \cite{zeitlin2004self}.
In recent years, the approach has given rise to structure-preserving integration methods for various spherical flows, including two-dimensional Euler flow \cite{cifani2023efficient}, Euler flow with transport noise \cite{ephrati2024exponential}, single- and multi-layer QG flow \cite{franken2024zeitlin, franken2025casimir}, thermal QG flow \cite{roop2025thermal}, and magnetohydrodynamics (MHD) \cite{modin2025spatio}.
Its conservation properties make the method well-suited for the numerical study of long-time dynamics and statistical flow features, such as the double cascade in two-dimensional turbulence \cite{cifani2022casimir, ephrati2025spectral}.
Here, we focus on the discretization for QG \cite{franken2024zeitlin}.
Since this is a non-standard method, we provide a brief outline below and refer interested readers to the cited sources. 
For a stand-alone introduction to Zeitlin's method and its mathematical background, see \cite{modin2026brief}.

In the absence of forcing and damping, both the Euler and QG systems are geometric \textit{Lie--Poisson} systems \cite{arnold1998topological, luesink2024geometric}.
The Lie--Poisson structure manifests through the conservation of energy, enstrophy, and higher-order integrated functions of PV (Casimirs).
The essential feature of Zeitlin's method is that the Lie--Poisson structure is retained even after discretization. Consequently, the discrete analogues of the energy, enstrophy, and integrated functions of PV are conserved numerically.
This remarkable feature is achieved via geometric quantization \cite{hoppe1989diffeomorphism, bordemann1994toeplitz, bordemann1991gl}, which relies on projecting smooth functions on the sphere to complex $N\times N$ skew-Hermitian matrices.
Here, $N$ is regarded as the resolution of the discretization.

Corresponding to the projection of smooth functions to skew-Hermitian matrices, the Laplace--Beltrami operator $\Delta$ is replaced by a discrete Laplacian $\Delta_N$.
Analogously, the spherical harmonics $Y_{l,m}$ are replaced by their discrete counterparts $T_{l,m}^N$, which are the eigenfunctions of the discrete Laplacian, \begin{equation}
    \Delta_N T_{l,m}^N = -l(l+1)T_{l,m}^N.
\label{eq:discrete_laplacian_eigenvalue}
\end{equation}
These \textit{matrix harmonics} form an orthonormal basis for the skew-Hermitian matrices.
Thus, the spectral properties of the Laplace--Beltrami operator are mimicked by the discrete Laplacian. 
Explicit expressions for $T_{lm}^N$ exist, however, these are expensive to compute. A more efficient approach is to compute these matrix harmonics by solving the eigenvalue problem \eqref{eq:discrete_laplacian_eigenvalue} directly \cite{cifani2023efficient}.

Discrete approximations to the PV $q$ are obtained by identifying spherical harmonics with matrix harmonics.
With the spherical harmonic coefficients of $q$ defined as $q_{l,m} = \langle q, Y_{l,m}\rangle$, the corresponding matrix approximation is \begin{equation}
    Q = \sum_{l=0}^{N-1}\sum_{m=-l}^l q_{l, m} T_{l,m}^N.
\end{equation}
Analogous expressions define the stream matrix $P$ and the topography matrix $H_\mathrm{topo}$.
This identification also allows the reconstruction of functions on the sphere, up to spherical harmonic degree $N$, from their matrix approximations.
In this representation, the Poisson bracket is replaced by a scaled matrix commutator, \begin{equation}
    \left\{\psi, q\right\} \to \frac{2}{\sqrt{N^2-1}}\left[P, Q\right],
\end{equation}
which approximates the bracket on the left up to an error $\mathcal{O}(N^{-2})$.

The appearance of $\gamma\mu^2\psi$ when applying the inhomogeneous Helmholtz operator requires a matrix representation for the product of two functions.
Because matrix multiplication is non-commutative, the projection from functions to matrices necessitates a symmetrized product. 
Letting $F$ and $G$ be the matrix representations of two functions $f$ and $g$, their product is represented as \begin{equation}
    fg \to -\frac{i}{2}\sqrt{\frac{N}{4\pi}}\left(FG + GF\right).
\end{equation}
The product on the left is approximated up to an error $\mathcal{O}(N^{-1})$ \cite{franken2024zeitlin}.
Recalling that $\mu=\sqrt{\frac{4\pi}{3}}Y_{1,0}$ and $\mu^2 = \frac{2}{3}\sqrt{\pi}Y_{0,0} + \frac{4}{3}\sqrt{\frac{\pi}{5}}Y_{2,0}$, we define their matrix analogues as $M = \sqrt{\frac{4\pi}{3}}T_{1,0}^N$ and $S=\frac{2}{3}\sqrt{\pi}T_{0,0}^N + \frac{4}{3}\sqrt{\frac{\pi}{5}}T_{2,0}^N$.

This leads to the Zeitlin QG equations, given by 
\begin{align}
    \label{eq:zeit_qg}
    \dot Q &+ \frac{2}{\sqrt{N^2-1}}[P,Q] = 0, \\
    Q&=\underbrace{\Delta_N P + \gamma \frac{i}{2}\sqrt{\frac{N}{4\pi}}\left(SP + PS\right)}_{H_N[P]} + \frac{2M}{\Ro} + \frac{i}{2}\sqrt{\frac{N}{4\pi}}\left(M H_\mathrm{topo} + H_\mathrm{topo}M\right)  \label{eq:QfromP}.
\end{align}
The expression for $H_N[P]$ is the discrete version of $H[\psi]$.
The matrix formulation of Eq. \eqref{eq:necessary_requirement} then becomes \begin{equation}
    \Delta_N P + \gamma \frac{i}{2}\sqrt{\frac{N}{4\pi}}\left(SP + PS\right) + \frac{2M}{\Ro} + \frac{i}{2}\sqrt{\frac{N}{4\pi}}\left(M H_\mathrm{topo} + H_\mathrm{topo}M\right) = \lambda P. 
    \label{eq:necessary_condition_zeitlin}
\end{equation}
The left-hand side of Eq. \eqref{eq:necessary_condition_zeitlin} is thus precisely the expression for $Q$, and the equation is compactly stated as $Q=\lambda P$.

In Zeitlin's model, integrals are replaced by matrix traces. 
Consequently, the energy and enstrophy are respectively given by \begin{align} \label{eq:zeitlin_energy}
    E_N[P] &= -\frac{1}{2N}\Tr(P^*H_N[P]), \\
    \mathcal{E}_N[P] &= \frac{1}{2N}\Tr(Q^* Q).
\end{align}

By solving Eq. \ref{eq:necessary_condition_zeitlin}, using the techniques in \cite{cifani2023efficient}, we obtain a mapping $\lambda \mapsto P$. 
From this $P$, we can compute the energy using Eq. \ref{eq:zeitlin_energy}, resulting in a mapping $f\colon \mathbb{R}\to \mathbb{R}$ which computes the energy for a given value of $\lambda$. 
Thus, finding the $\lambda$ that yields a specified target energy $E_*$ is a root-finding problem. 
We solve this problem using the bisection method as follows.

The bisection method requires an interval $[\lambda_-, \lambda_+]$, where $f(\lambda_-)-E_*$ and $f(\lambda_+)-E_*$ have opposite signs. 
We select $\lambda_+$ to be sufficiently large so that $f(\lambda_+)<E_*$. In practice, we select $\lambda_+ = 10^6$ as this results in a near-zero energy. 
We denote by $\omega_0^N$ the largest eigenvalue of $H_N$. As $\lambda \to \omega_0^N$, the energy goes to infinity. 
Computing $\omega_0^N$ is possible by exploiting the block-diagonal structure of the Zeitlin-operator $H_N$.
As noted in \cite{cifani2023efficient}, the operator decomposes into $(2N-1)$ tridiagonal blocks $H_N^m$ of size $N-|m|$.
The eigenvalues of $H_n$ are thus given as the union of the eigenvalues of each block which are computed using the SSTEBZ method (see \cite{kahan1966accurate}).

Our initial guess of the lower endpoint $\lambda_-$ is $\omega_0^N + \delta$, where $\delta$ is a small offset. If $f(\omega_0^N + \delta)>E_*$, we take this as our lower end point. Otherwise, we consider $\omega_0^N+\delta/2$ and repeat the procedure if $f(\omega_0^N+\delta/2) < E_*$, until the left end point produces an energy larger than $E_*$. In practice, we set $\delta = 10^{-2}$. 
The bisection algorithm then computes the midpoint $\lambda_\mathrm{mid} = (\lambda_+ + \lambda_-)/2$ and the corresponding value $f(\lambda_\mathrm{mid})$. If $f(\lambda_\mathrm{mid})>E_*$, then the algorithm repeats with the interval $[\lambda_\mathrm{mid},\lambda_+]$, otherwise it repeats with $[\lambda_-,\lambda_\mathrm{mid}]$.
When the interval width is sufficiently small, the midpoint is returned as the root. 
In practice, we require the interval $[a,b]$ to satisfy the condition $|b-a| < \varepsilon_1 + |b|\varepsilon_2$, where $\varepsilon_1 = 2\times 10^{-12}$ and $\varepsilon_2 \approx 8.9\times 10^{-16}$.

\section{Numerical demonstrations}\label{sec:numerical_demonstrations}
In this section, we numerically compute minimum-enstrophy solutions using Zeitlin’s method as highlighted in  \Cref{sec:computational_method}. 
In what follows, we adopt a resolution of $N=1024$ (see Appendix \ref{app:matrix_size}).
A parameter study is carried out first, to demonstrate the qualitative features of these solutions in the different limiting regimes laid out in  \Cref{subsec:asymptotic_regimes}.
Subsequently, we study minimum-enstrophy solutions using the parameters for the Jovian atmosphere in  \Cref{tab:physical_parameters}.

\subsection{Parameter study}
\begin{figure}[ht]
    \centering
    \includegraphics[width=0.7\linewidth]{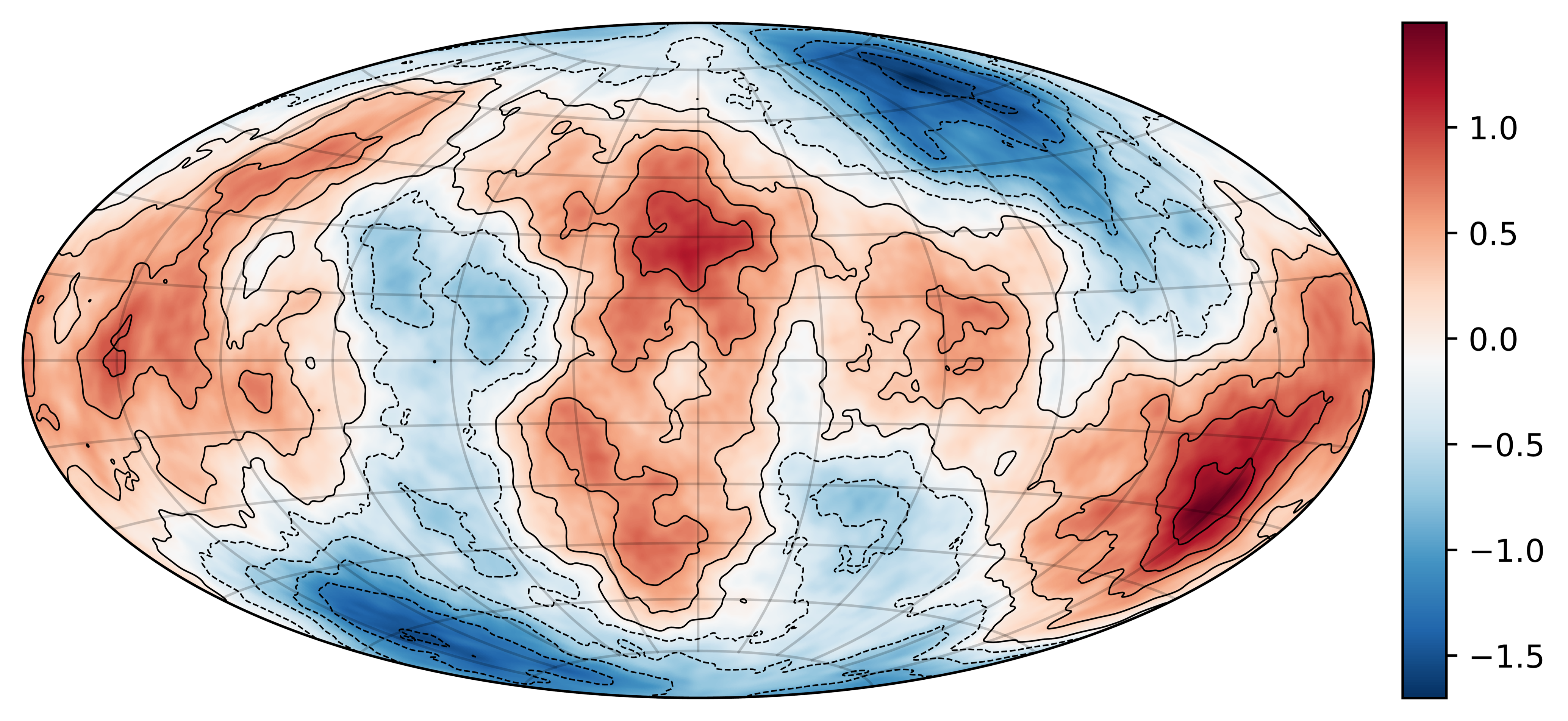}
    \caption{Topography used throughout the reported test cases.}
    \label{fig:topography}
\end{figure}

In line with the discussion on the asymptotic limits in \Cref{subsec:asymptotic_regimes}, we carry out several parameter sweeps and assess the corresponding outcomes.
The adopted parameters do not necessarily resemble physical regimes, but are used to verify the simplified settings of  \Cref{subsec:asymptotic_regimes}. 
In particular, we separately vary the Rossby number, Lamb’s parameter $\gamma$, and the energy (or, equivalently, the Lagrange multiplier $\lambda$), to assess their effects on the computed minimum-enstrophy solutions.
Throughout the reported tests, we use a single, fixed topography which is a randomly generated element in  $\mathfrak{su}(1024)$ ($1024\times 1024$ skew-Hermitian matrices with zero trace) by 
\begin{align}
    \label{eq:topo}
    H_\mathrm{topo} = \sum_{l=2}^{127} \sum_{m=-l}^l a_{l,m} \frac{1}{(l^2+l+m+1)^\alpha} T_{l,m}^N
\end{align}
where $\alpha = 1.0$ and $a_{l,m}$  are independent, identically distributed standard normal random variables. Note that $l^2+l+m+1$ is the index of coefficient $(l,m)$ in the linear ordering of the spherical harmonics. 
The topography profile used is throughout, unless otherwise stated, and is shown in Figure \ref{fig:topography}. 

\subsubsection{Varying the Rossby number}
\begin{figure}[ht]
    \centering
    \includegraphics[width=0.9\linewidth]{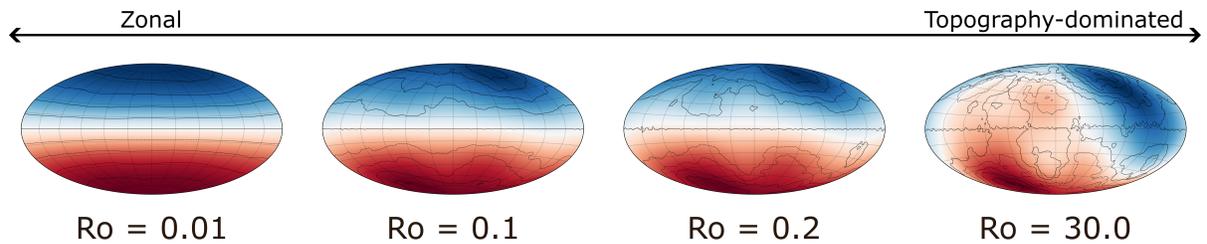}
    \caption{Qualitative comparison of minimum-enstrophy solutions at different Rossby numbers. The colored background indicates the value of the stream function, the contour lines depict equispaced values of $\mu h - 2\mu/\Ro$. The limits of the color maps are specific to each panel to visualize the qualitative differences. For small Rossby numbers, rotation effects dominate and the minimum-enstrophy solutions are zonal. For increasing Rossby numbers, topographic effects become more pronounced and cause the flow to be topographically trapped, particularly visible near the poles. }
    \label{fig:ro-variation}
\end{figure}

As a first test, we vary the Rossby number. 
The results are shown in \Cref{fig:ro-variation}, where the colored background indicates the value of the stream function and the contour lines depict equispaced values of $\mu h - 2\mu/\Ro$.
Here, we choose $\gamma = 5$.
In general, the stream function is found to be proportional to $\mu h - 2\mu/\Ro$ for the  Rossby numbers .
These observations align with the predictions of \Cref{subsec:asymptotic_regimes}.
Namely, $\mu h - 2\mu/\Ro$ is dominated by the rotational term for small Rossby numbers, causing topography effects to decrease, whereas the opposite effect is observed for large Rossby numbers.
This is particularly visible from the contour lines in  \Cref{fig:ro-variation}.
Consequently, the flow becomes increasingly dominated by topography at high Rossby numbers.
This is best seen in the right-most panel, where coherent structures form in the stream function at locations where topographical features exist. 
At smaller Rossby numbers, such topographical trapping is only observed closer to the poles.

\subsubsection{Varying Lamb’s parameter}
We continue by varying Lamb’s parameter $\gamma$ while keeping the Rossby number fixed at $\Ro = 10$.
In this case, an increase of $\gamma$ corresponds to a decrease of the average fluid depth.
Due to the appearance of $\gamma$ in the denominator of equation \eqref{eq:large_lamb_limit_1}, the stream function is expected to decrease in magnitude for increasing $\gamma$.
We therefore focus on both the qualitative features of $\psi$ and its magnitude.

\begin{figure}[ht]
     \centering
     \begin{subfigure}[b]{0.3\textwidth}
         \centering
         \includegraphics[width=\textwidth]{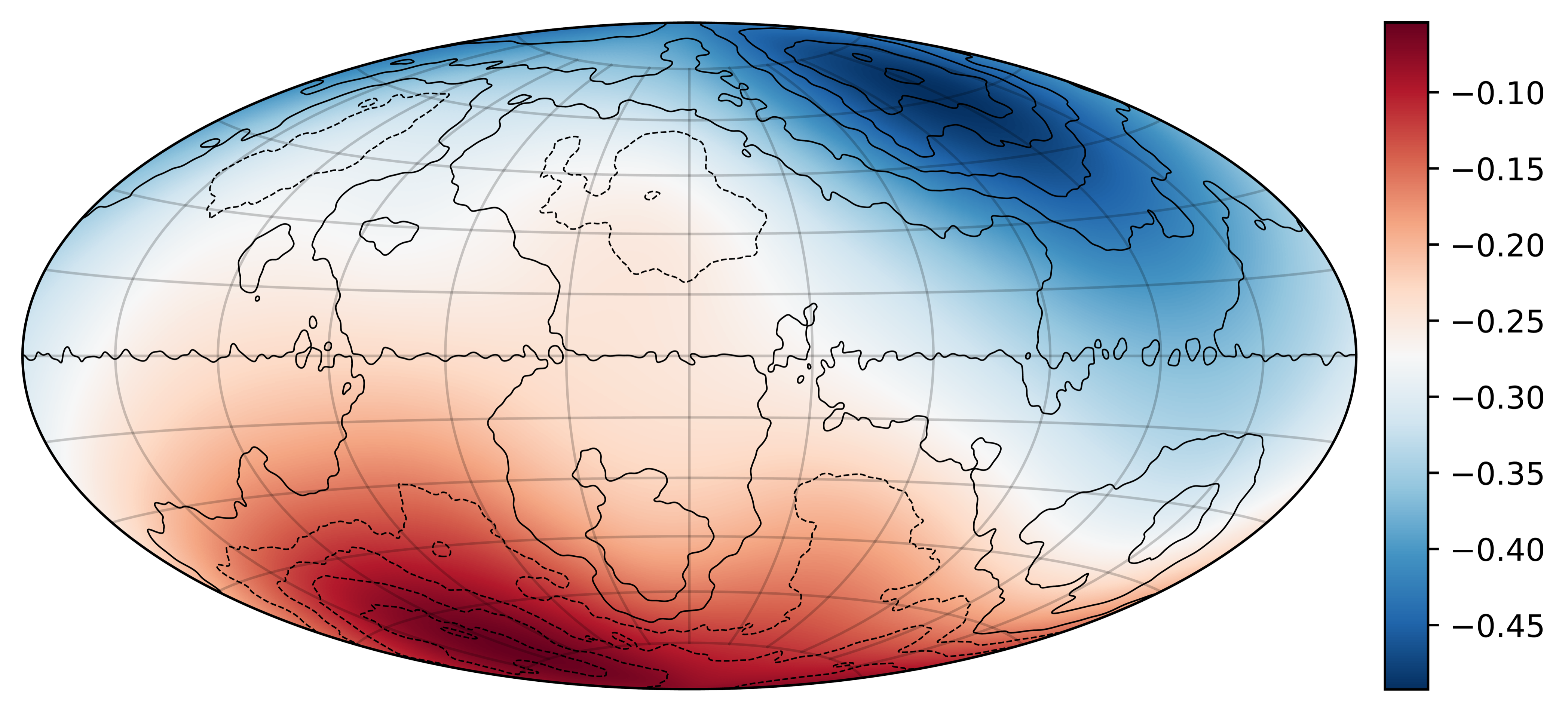}
         \caption{$\gamma = 0.5$}
         \label{fig:gammalow}
     \end{subfigure}
     \hfill
     \begin{subfigure}[b]{0.3\textwidth}
         \centering
         \includegraphics[width=\textwidth]{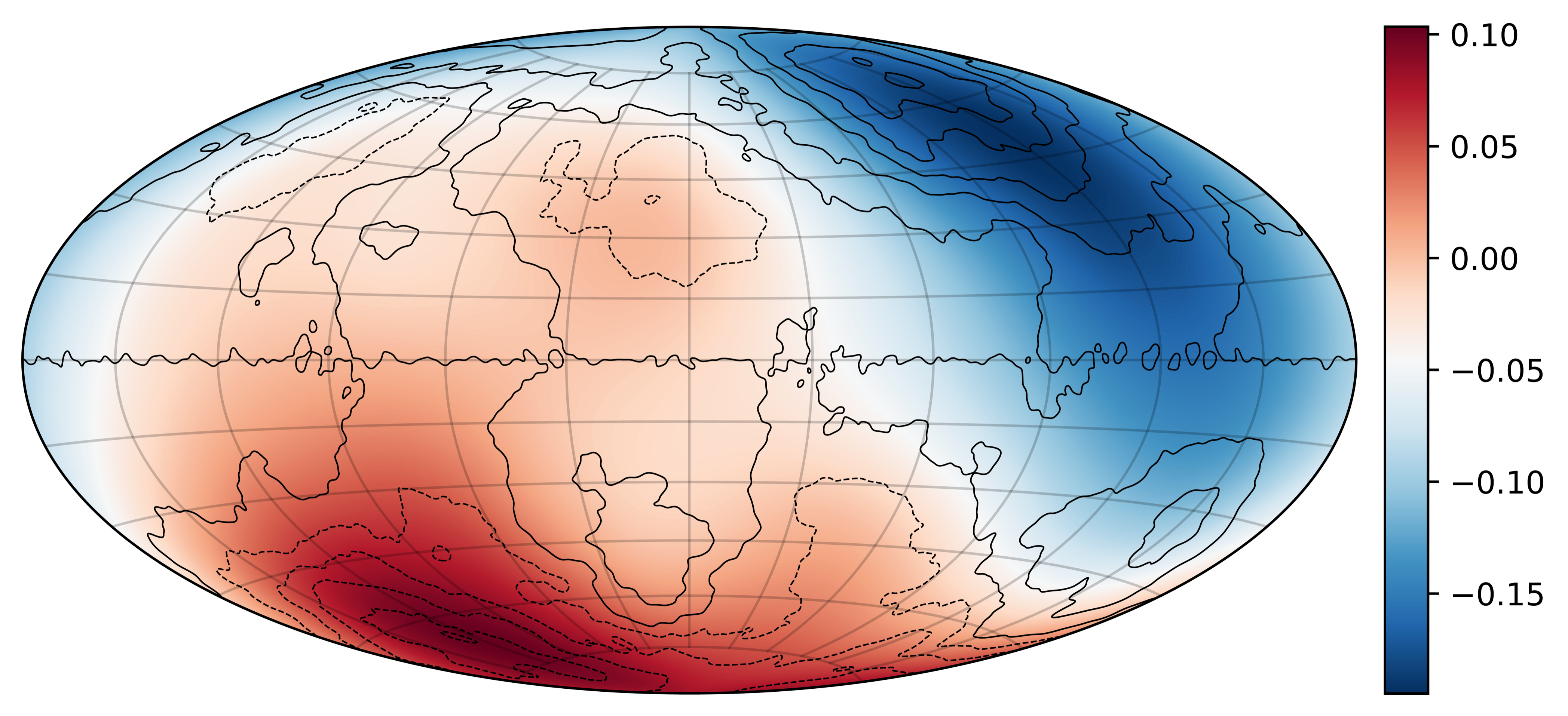}
         \caption{$\gamma = 5.0$}
         \label{fig:gammamid}
     \end{subfigure}
     \hfill
     \begin{subfigure}[b]{0.3\textwidth}
         \centering
         \includegraphics[width=\textwidth]{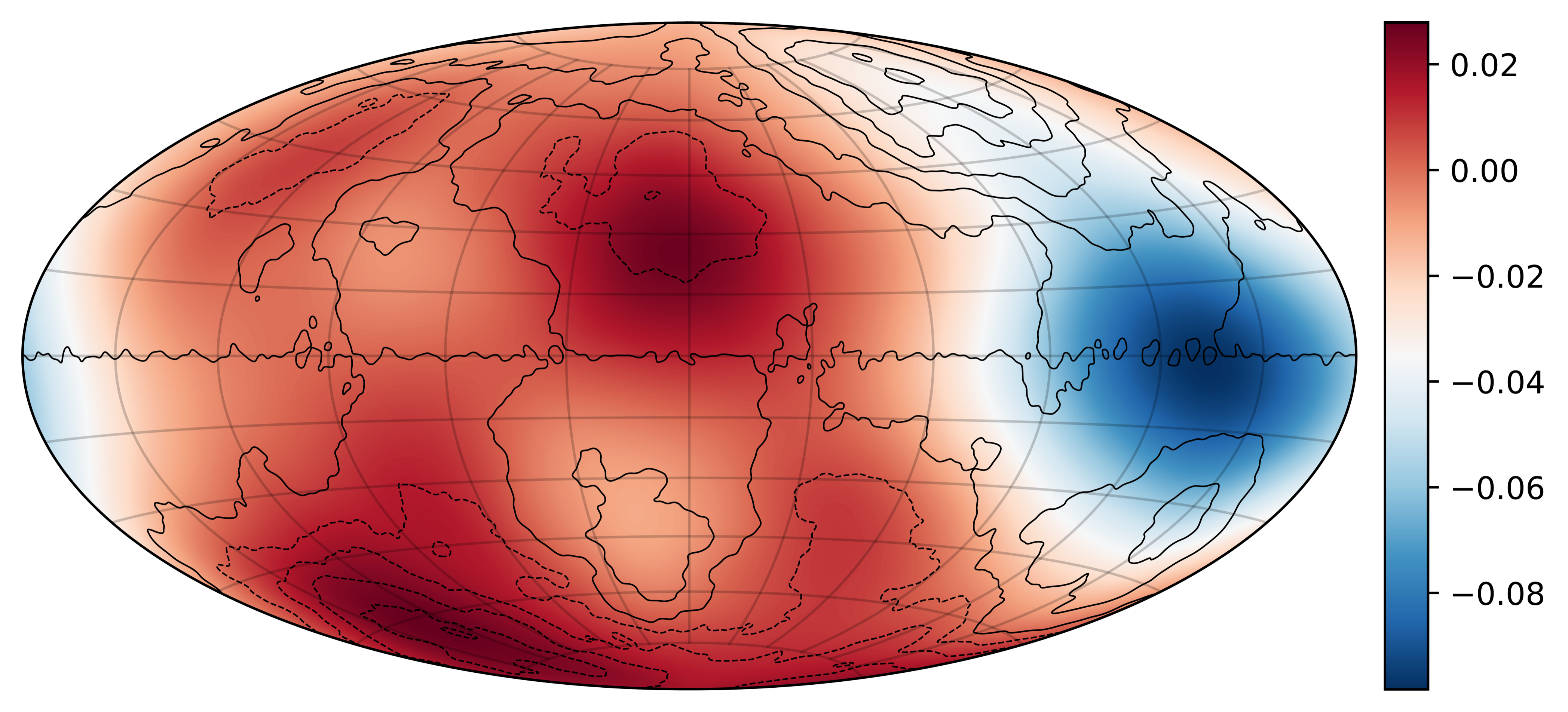}
         \caption{$\gamma = 50.0$}
         \label{fig:gammahigh}
     \end{subfigure}
        \caption{Qualitative comparison of minimum-enstrophy solutions for different values of Lamb's parameter $\gamma$ at fixed values of the Rossby number. The colored background indicates the value of the stream function, the contour lines depict equispaced values of $\mu h - 2\mu/\Ro$. As $\gamma$ increases, the magnitude of the stream function is suppressed and non-zonal circulation near the equator is enhanced.}
        \label{fig:gammavar}
\end{figure}

The results are shown in \Cref{fig:gammavar}.
The left panel (low $\gamma$) shows that the flow follows the topographic contour lines.
As $\gamma$ increases, however, a qualitative shift in the stream function emerges, even though the contours of $\mu h - 2\mu/\Ro$ remain unchanged.
This behaviour reflects a balance between the two competing effects, being topographical trapping of the flow and enhanced non-zonal circulation near the equator.
Although increasing $\gamma$ generally suppresses the stream function, this tendency is offset near the equator as $\mu$ approaches zero, allowing a local strengthening of the circulation.

\subsubsection{Varying the energy}
\begin{figure}[ht]
     \centering
     \begin{subfigure}[b]{0.3\textwidth}
         \centering
         \includegraphics[width=\textwidth]{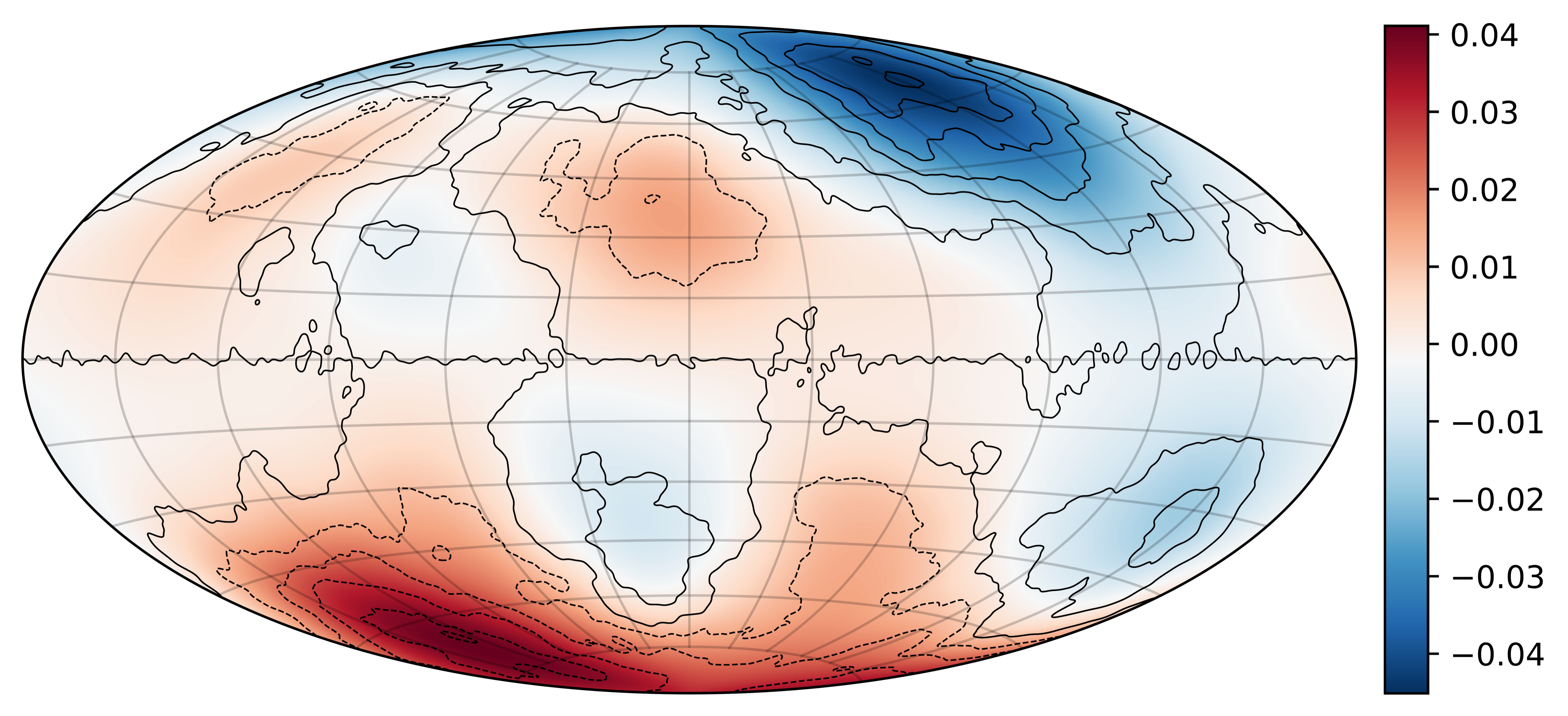}
         \caption{ $E = 0.001$}
         \label{fig:energylow}
     \end{subfigure}
     \hfill
     \begin{subfigure}[b]{0.3\textwidth}
         \centering
         \includegraphics[width=\textwidth]{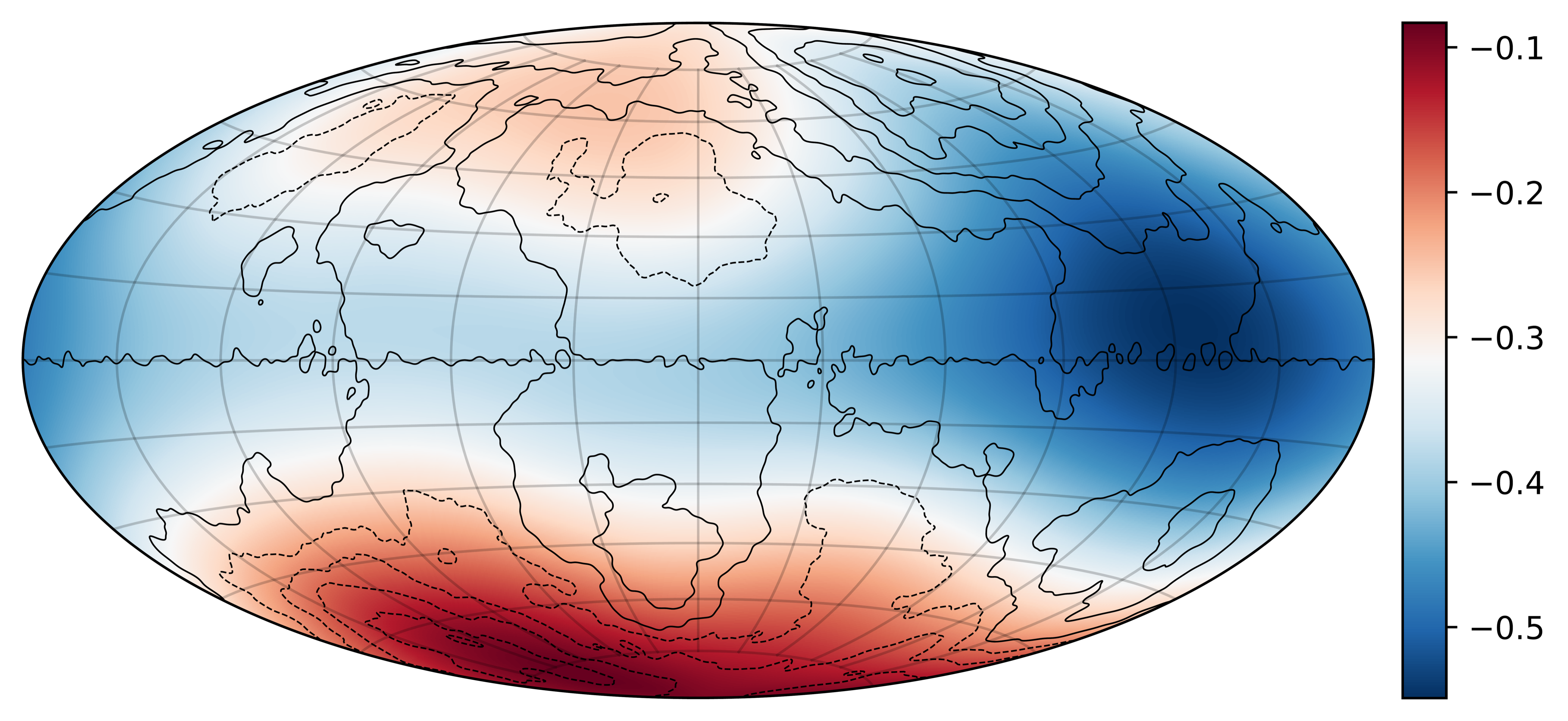}
         \caption{ $E = 0.1$}
         \label{fig:energymid}
     \end{subfigure}
     \hfill
     \begin{subfigure}[b]{0.3\textwidth}
         \centering
         \includegraphics[width=\textwidth]{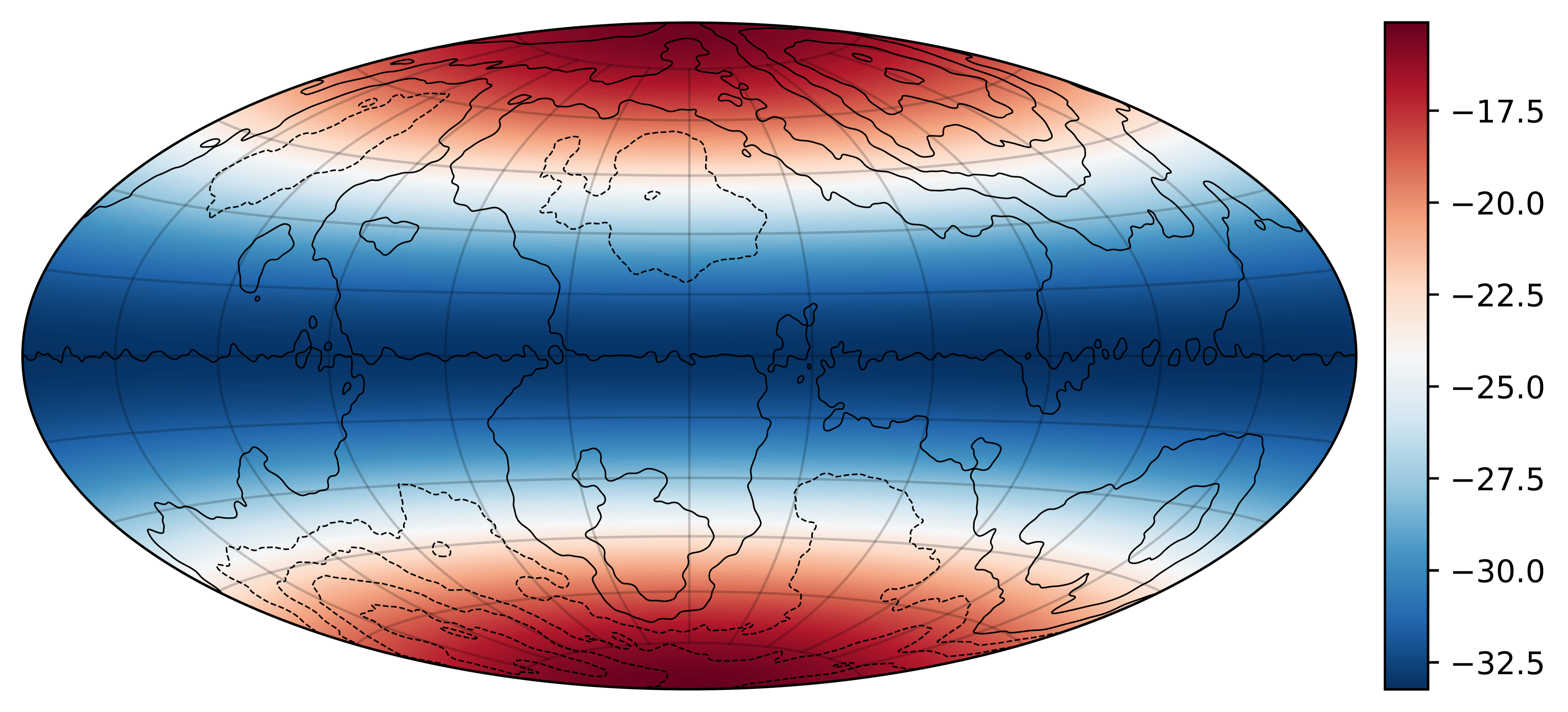}
         \caption{$ E = 500$}
         \label{fig:energyhigh}
     \end{subfigure}
        \caption{Qualitative comparison of minimum-enstrophy solutions for different energy values, at fixed Rossby number and Lamb's parameter $\gamma$. The colored background indicates the value of the stream function, the contour lines depict equispaced values of $\mu h - 2\mu/\Ro$. As the energy increases (from left to right), the solutions are less influenced by the topography and instead develop a zonal structure.}
        \label{fig:energyvar}
\end{figure}

Naturally, the energy of the flow determines the corresponding minimum-enstrophy solution.
On the torus, lower energies lead to vortices locked to topographic features \cite{siegelman2023two} and, following \Cref{subsec:asymptotic_regimes}, the same is expected for solutions on the sphere.
This is illustrated in \Cref{fig:energyvar}, showing the attained minimum-enstrophy solutions at various energy levels where $\gamma = 5$ and $\Ro = 10$. 
At low energy, the stream function is found to follow the topography throughout the entire domain.
At the intermediate energy level, the solution is seen to develop a zonal structure near the equator, while some localization of the stream function near topographic features is observed at higher latitudes.
Finally, at sufficiently high energy, the resulting stream function appears completely zonal and hardly affected by the topography.

\subsection{Physical parameters: Jovian atmosphere}
We proceed by computing the minimum-enstrophy solution using the parameters for Jupiter's atmosphere as reported in Table \ref{tab:physical_parameters}. 
This provides an example for the solution behaviour using physical parameters.
Specifically, we adopt $\Ro = 8.9\times 10^{-3}$, $\gamma = 1.3 \times 10^2$, $N = 1024$, and use the topography in \Cref{fig:topography}. 
The energy level is set to $E_* = 1.67$, resulting in $\lambda_* = -0.0154$.  

\begin{figure}[ht]
    \centering
    \includegraphics[width=0.8\linewidth]{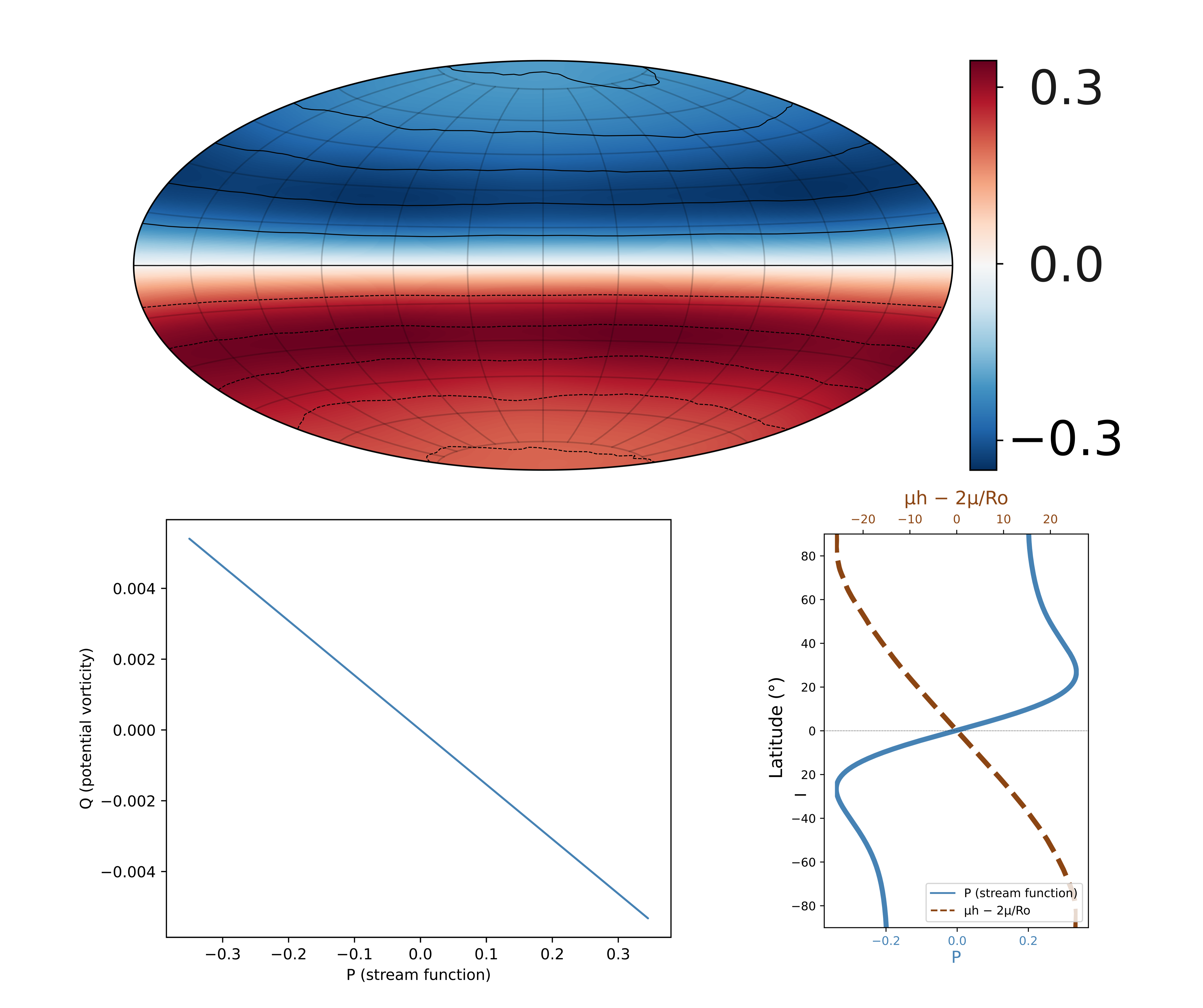}
    \caption{Minimum-enstrophy solution using the parameters of the Jovian atmosphere (see Table \ref{tab:physical_parameters}). In the top panel, the colored background indicates the value of the stream function, the contour lines depict equispaced values of $\mu h -2\mu/\Ro$.
    The stream function is predominantly zonal, attributed to the small Rossby number, which also causes $\mu h -2\mu/\Ro$ to be affected by the topography only near the poles.
    The bottom left panel displays the linear relation between pointwise values of the PV and the stream function.
    The bottom right panel shows the profiles of the stream function and $\mu h - 2\mu/\Ro$ along the central meridian of the figure in the top panel.}
    \label{fig:jupiter}
\end{figure}

The results are shown in Figure \ref{fig:jupiter}.
In this strongly rotating regime, the influence of the topography is observed to be negligible.
The resulting stream function is found to be zonal.
The contour lines of $\mu h - 2\mu/\Ro$ are almost parallel to the equator, and only near the poles are they observed to be influenced by the topography. 
This is in line with the previously presented results for small Rossby numbers.

The bottom left panel of Figure \ref{fig:jupiter} displays the relation between pointwise values of the obtained potential vorticity and the stream function, confirming the linear relation $Q=\lambda P$ arising from Equation \eqref{eq:necessary_condition_zeitlin} is satisfied.
The bottom right panel shows the profiles of the stream function and the function $\mu h - 2\mu/\Ro$ along the central meridian of the top panel.
Here, we observe an alignment of the stream function with $\mu h - 2\mu/\Ro$ near the equator.

The minimum-enstrophy solutions corresponding to the parameter sets for atmospheric and oceanic flows on Earth, as presented in Table \ref{tab:physical_parameters}, have also been computed.
However, no significant differences were found compared to the results in Figure \ref{fig:jupiter}, and these are therefore not further elaborated upon for the sake of brevity.

\subsection{Structure-preserving simulation of perturbations}
In the following numerical tests, we assess the stability of the  minimum-enstrophy solutions.
The theory in \Cref{sec:minens} states that the minimum-enstrophy stream function $P_*$ provides a stable equilibrium of \Cref{eq:zeit_qg}. 
To verify this numerically, we compute a solution as described in Section \ref{sec:computational_method}, subsequently perturb the solution, and establish that the solution norm remains bounded when evolving according to the QG dynamics.

In these tests, we set $N = 128$, $\Ro=10$, $\gamma=5$ and use the topography depicted in \Cref{fig:topography}. 
We then compute $Q_*$ from $P_*$ using Eq. \eqref{eq:QfromP} and perturb it by adding a random element in $\mathfrak{su}(N)$ ($N\times N$ skew-Hermitian matrices with zero trace; corresponding to a zero-mean perturbation), scaled such that the enstrophy of the perturbed state is given by $(1+\epsilon)\mathcal{E}[P_*]$. 
Here, we choose $\epsilon = 0.01, 0.05, 0.1$ and $0.2$.

Using this perturbed state as initial condition, Eq. \eqref{eq:zeit_qg} is integrated for a total of 500 time units.
The isospectral midpoint method (IsoMP) \cite{modin2020casimir} is adopted for time integration, since it numerically preserves enstrophy and nearly preserves energy.
Here, we use a time step size $\mathrm{d}t=0.5/\sqrt{N^2-1}$.
A total of 100 perturbations are independently and randomly computed per value of $\epsilon$, resulting in an ensemble of independent simulations.

\begin{figure}[H]
    \centering
    \begin{subfigure}[t]{0.4\linewidth}
        \includegraphics[width=\linewidth]{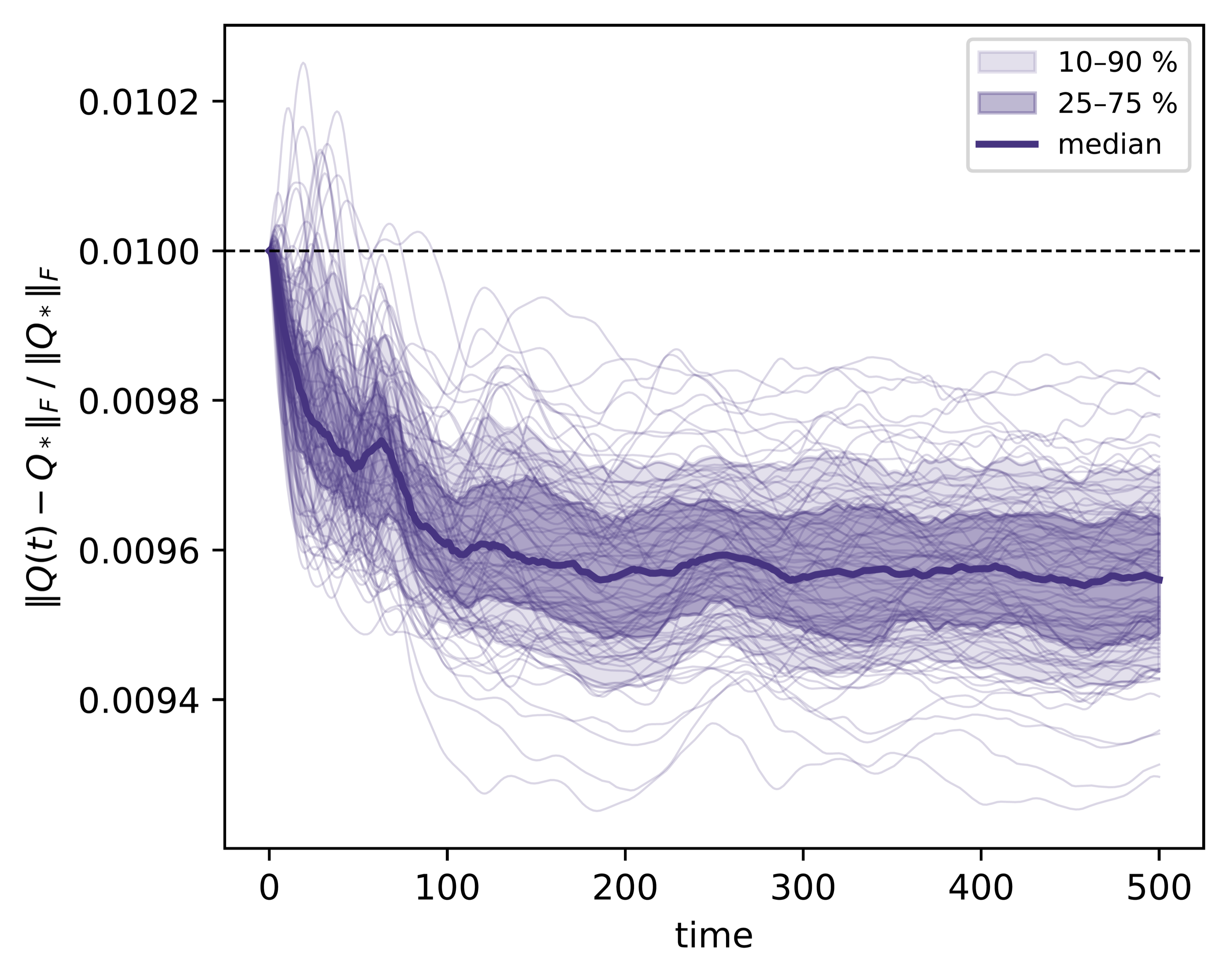}
        \caption{$\epsilon = 0.01$}
    \end{subfigure}
    \hfill
    \begin{subfigure}[t]{0.4\linewidth}
         \includegraphics[width=\linewidth]{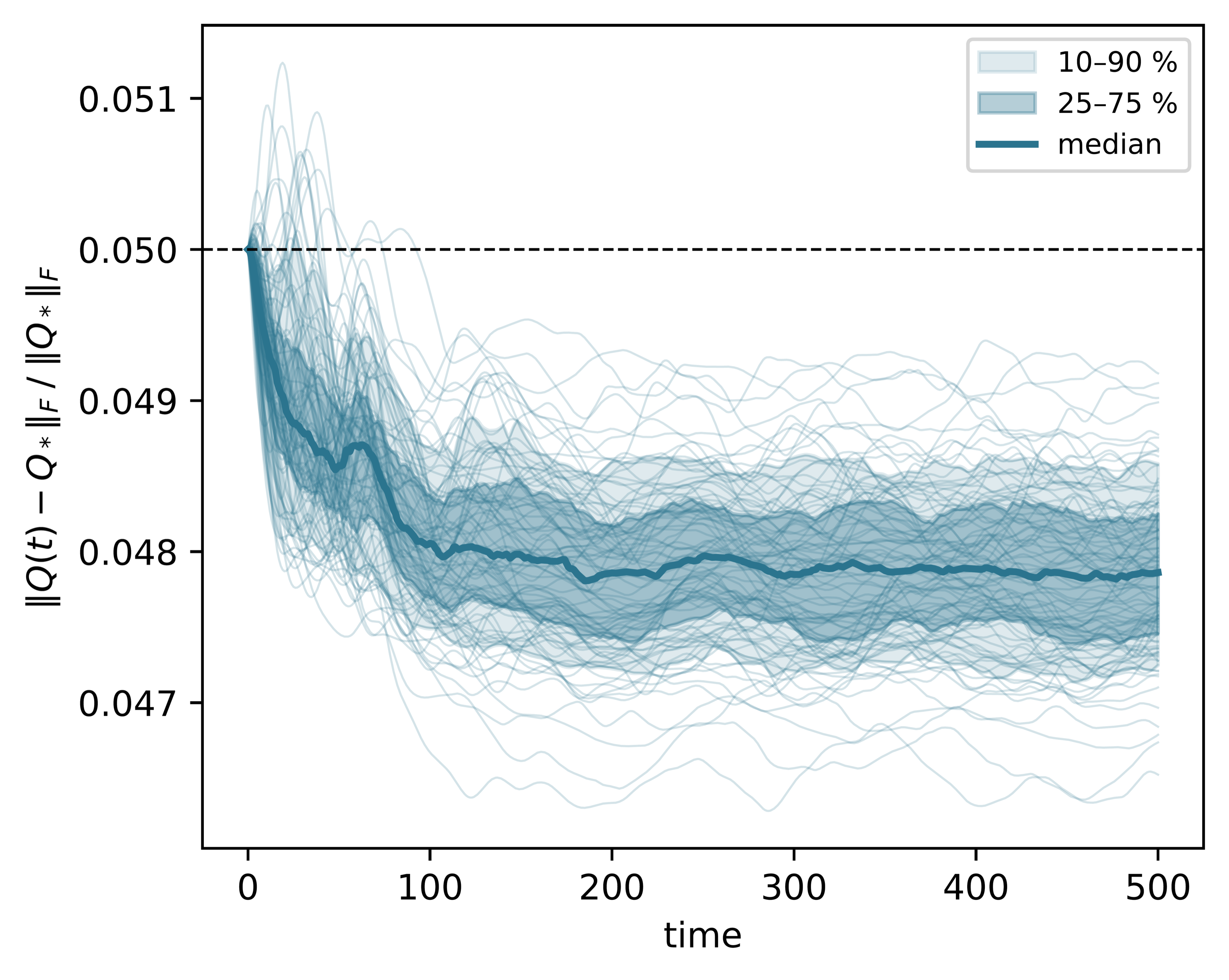}
        \caption{$\epsilon = 0.05$}
    \end{subfigure}
    \\[0.5em]
    \begin{subfigure}[t]{0.4\linewidth}
         \includegraphics[width=\linewidth]{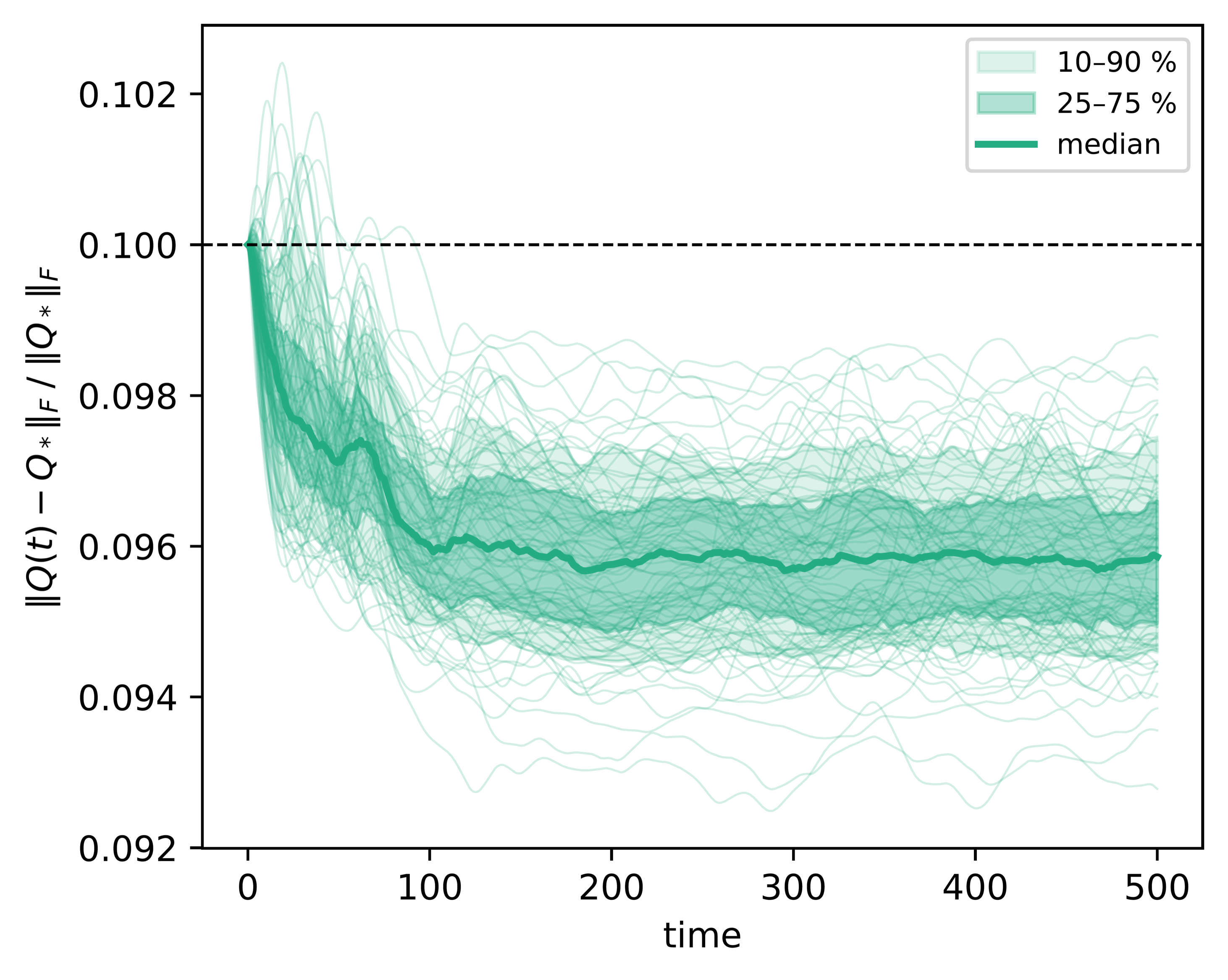}
        \caption{$\epsilon = 0.1$}
    \end{subfigure}
    \hfill
    \begin{subfigure}[t]{0.4\linewidth}
        \includegraphics[width=\linewidth]{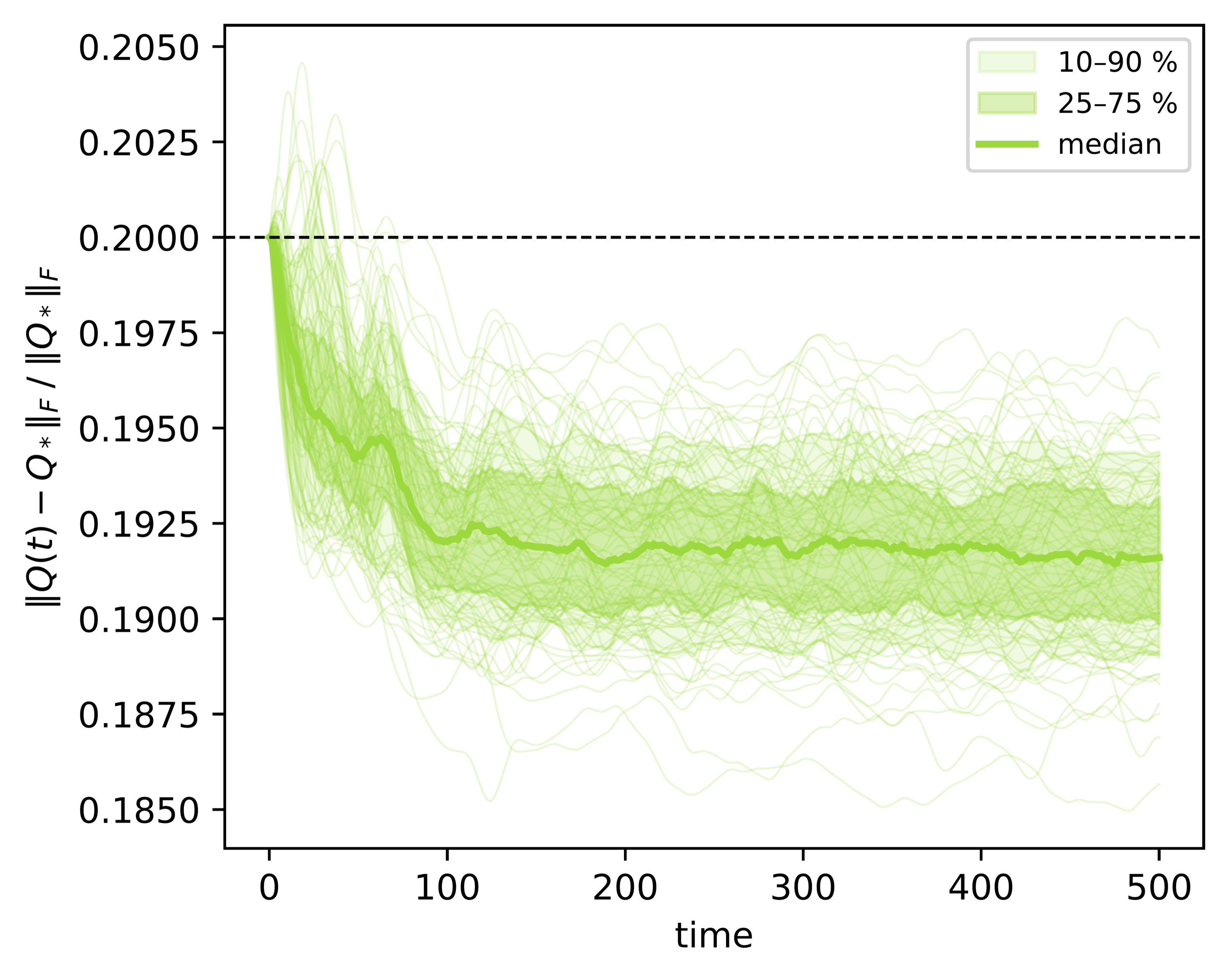}
        \caption{$\epsilon= 0.2$}
    \end{subfigure}
    \caption{Evolution of the relative deviation from the equilibrium $Q_*$ over time, for perturbation magnitudes $\epsilon \in \{0.01, 0.05, 0.1, 0.2\}$.
    For each $\epsilon$, 100 random perturbations are independently computed, yielding a simulation ensemble of which the median and the $10^\mathrm{th}$–$90^\mathrm{th}$ and $25^\mathrm{th}$–$75^\mathrm{th}$ percentile bands are shown.}
    \label{fig:stability_time}
\end{figure}

\begin{figure}[H]
    \centering
    \includegraphics[width=0.4\linewidth]{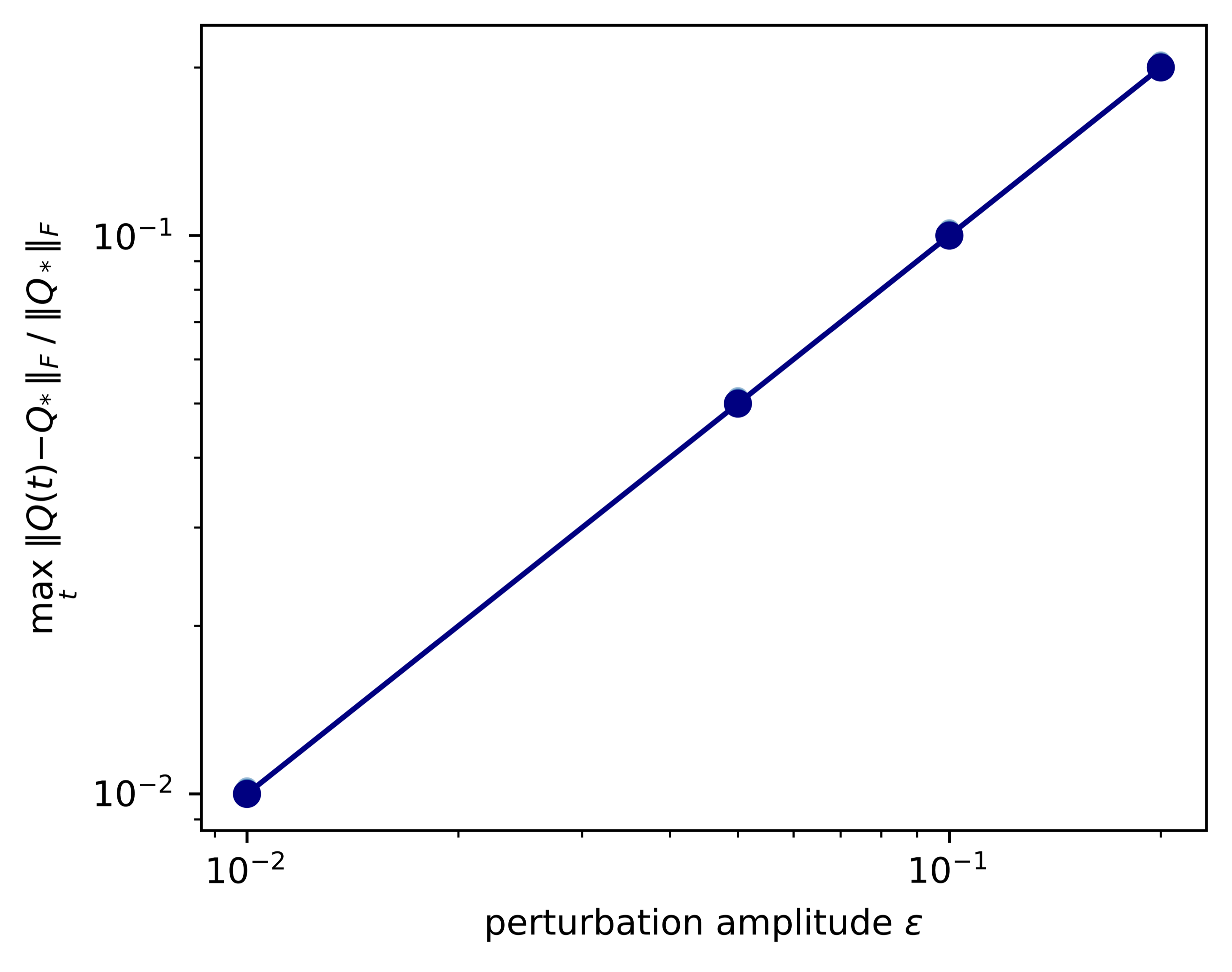}
    \caption{The maximum relative deviation from the equilibrium $Q_*$ displayed as a function of the perturbation magnitude $\epsilon$.}
    \label{fig:stability_linear}
\end{figure}

The evolution of the relative deviation from $Q_*$ is shown in \Cref{fig:stability_time} along with the median and the $10^\mathrm{th}$–$90^\mathrm{th}$ and $25^\mathrm{th}$–$75^\mathrm{th}$ percentile bands.
For all values of $\epsilon$, the relative deviation remains bounded and a qualitatively similar evolution of the relative deviation is observed. 
Correspondingly, the relation between the maximum measured relative deviation and $\epsilon$ is linear, as depicted in \Cref{fig:stability_linear}.
That is, the relative deviation is of order $\epsilon$. 
These results are consistent with Lyapunov stability of the equilibrium.

\section{Conclusions and outlook}\label{sec:conclusions}

In this paper, we have extended the selective decay hypothesis of Bretherton and Haidvogel to the spherical setting, incorporating the effects of curvature and rotation into the study of minimum-enstrophy solutions of flow over topography.
By analysing the spherical quasi-geostrophic equations, we derived the corresponding Euler--Lagrange equations, which characterise the minimum-enstrophy solutions as flow equilibria where the potential vorticity and the stream function are linearly related.
A fundamental distinction from previous work is the inclusion of the full nonlinear Coriolis term, resulting in the presence of an inhomogeneous Helmholtz operator instead of a Laplacian.
Using the spectral properties of this operator, we provided a formal proof of existence and nonlinear stability of the minimum-enstrophy solutions.

Three simplifying asymptotic regimes of solutions were identified through a spectral expansion. 
These regimes indicate that the spherical geometry and rotation introduce competing physical effects with a pronounced latitudinal dependency.
At lower energy levels or lower rotation rates, the flow tends to align with the topography near the poles, whereas zonal flow patterns emerge naturally near the equator. 
The balance of these effects is determined by the energy, rotation rate, and Lamb's parameter.
These analytic predictions were corroborated by numerical computations using Zeitlin's self-consistent truncation for ideal flows on the sphere.
The numerical results confirmed the linear relationship between the potential vorticity and the stream function and that the solutions remain stable under perturbations.

Several avenues for future research can be explored based on the currently presented work.
Here, we have shown that many parallels can be drawn between spherical solutions and those in planar approximations.
Future work could investigate the spherical analogues of recently performed analysis on the torus.
For example, the critical energy level that separates vorticity mixing from topographical confinement \cite{siegelman2023two} and its relation to Lamb's parameter, the Rossby number and the latitude can be investigated.
Alternatively, of interest is the existence of a `condensate branch' of solutions \cite{gallet2024two} on the sphere, where a large-scale condensate emerges along with features at the scale of the topography.
A numerical investigation of topography roughness may additionally be carried out for comparison against flows on the torus \cite{priya2026two}.
Such extensions would further bridge the gap between idealized two-dimensional turbulence theory and multi-scale dynamics of planetary atmospheres and oceans.

\section*{Acknowledgments}
We are grateful to Klas Modin and Michael Roop for insightful discussions during the preparations of this work.
SE was supported in part by the Swedish Research Council (VR) through grant no. 2022-03453 and in part by the US AFOSR grant FA8655-25-1-7465, Stochastic Plasma Physics Dynamics (SPPD).
EJ was funded by Knut and Alice Wallenberg Foundation grant no. 2024.0440. 
The computations  were enabled by resources provided by Chalmers e-Commons at Chalmers and  by resources provided by the National Academic Infrastructure for Supercomputing in Sweden (NAISS), partially funded by the Swedish Research Council through grant agreement no. 2022-06725.
\bibliographystyle{abbrv}
\bibliography{references}

\begin{thebibliography}{10}

\bibitem{arnold1966priori}
V.~I. Arnold.
\newblock An a priori estimate in the theory of hydrodynamic stability.
\newblock {\em Izv. Vyssh. Uchebn. Zaved. Mat.[Sov. Math. J.]}, 5:3, 1966.

\bibitem{arnold1998topological}
V.~I. Arnold and B.~A. Khesin.
\newblock {\em Topological methods in hydrodynamics}.
\newblock Springer, 1998.

\bibitem{biskamp2001two}
D.~Biskamp and E.~Schwarz.
\newblock On two-dimensional magnetohydrodynamic turbulence.
\newblock {\em Physics of Plasmas}, 8(7):3282--3292, 2001.

\bibitem{boffetta2012two}
G.~Boffetta and R.~E. Ecke.
\newblock Two-dimensional turbulence.
\newblock {\em Annual review of fluid mechanics}, 44(1):427--451, 2012.

\bibitem{bordemann1991gl}
M.~Bordemann, J.~Hoppe, P.~Schaller, and M.~Schlichenmaier.
\newblock {$\mathfrak{gl} (\infty$)} and geometric quantization.
\newblock {\em Communications in Mathematical Physics}, 138(2):209--244, 1991.

\bibitem{bordemann1994toeplitz}
M.~Bordemann, E.~Meinrenken, and M.~Schlichenmaier.
\newblock Toeplitz quantization of {K}{\"a}hler manifolds and {$\mathfrak{gl} (N), N\to\infty$} limits.
\newblock {\em Communications in Mathematical Physics}, 165(2):281--296, 1994.

\bibitem{brands1999maximum}
H.~Brands, P.~Chavanis, R.~Pasmanter, and J.~Sommeria.
\newblock Maximum entropy versus minimum enstrophy vortices.
\newblock {\em Physics of Fluids}, 11(11):3465--3477, 1999.

\bibitem{bretherton1976two}
F.~P. Bretherton and D.~B. Haidvogel.
\newblock Two-dimensional turbulence above topography.
\newblock {\em Journal of Fluid Mechanics}, 78(1):129--154, 1976.

\bibitem{carnevale1987nonlinear}
G.~F. Carnevale and J.~S. Frederiksen.
\newblock Nonlinear stability and statistical mechanics of flow over topography.
\newblock {\em Journal of Fluid Mechanics}, 175:157--181, 1987.

\bibitem{charney1962stability}
J.~G. Charney and M.~E. Stern.
\newblock On the stability of internal baroclinic jets in a rotating atmosphere.
\newblock {\em Journal of Atmospheric Sciences}, 19(2):159--172, 1962.

\bibitem{cifani2022casimir}
P.~Cifani, M.~Viviani, E.~Luesink, K.~Modin, and B.~J. Geurts.
\newblock Casimir preserving spectrum of two-dimensional turbulence.
\newblock {\em Physical Review Fluids}, 7(8):L082601, 2022.

\bibitem{cifani2023efficient}
P.~Cifani, M.~Viviani, and K.~Modin.
\newblock An efficient geometric method for incompressible hydrodynamics on the sphere.
\newblock {\em Journal of Computational Physics}, 473:111772, 2023.

\bibitem{cressman1958barotropic}
G.~P. Cressman.
\newblock Barotropic divergence and very long atmospheric waves.
\newblock {\em Monthly weather review}, 86(8):293--297, 1958.

\bibitem{daley1983linear}
R.~Daley.
\newblock Linear non-divergent mass-wind laws on the sphere.
\newblock {\em Tellus A: Dynamic Meteorology and Oceanography}, 35(1):17--27, 1983.

\bibitem{ephrati2024exponential}
S.~Ephrati, E.~Jansson, A.~Lang, and E.~Luesink.
\newblock An exponential map free implicit midpoint method for stochastic {Lie--Poisson} systems.
\newblock {\em arXiv preprint arXiv:2408.16701}, 2024.

\bibitem{ephrati2025spectral}
S.~Ephrati, E.~Jansson, and K.~Modin.
\newblock On spectral scaling laws for averaged turbulence on the sphere.
\newblock {\em Physica D: Nonlinear Phenomena}, page 134808, 2025.

\bibitem{franken2024critical}
A.~Franken, E.~Luesink, S.~Ephrati, and B.~Geurts.
\newblock Critical latitude in global quasi-geostrophic flow on a rotating sphere.
\newblock {\em arXiv preprint arXiv:2409.05432}, 2024.

\bibitem{franken2024zeitlin}
A.~D. Franken, M.~Caliaro, P.~Cifani, and B.~J. Geurts.
\newblock Zeitlin truncation of a shallow water quasi-geostrophic model for planetary flow.
\newblock {\em Journal of Advances in Modeling Earth Systems}, 16(6):e2023MS003901, 2024.

\bibitem{franken2025casimir}
A.~D. Franken, E.~Luesink, S.~R. Ephrati, and B.~J. Geurts.
\newblock Casimir preserving numerical method for global multi-layer quasi-geostrophic turbulence.
\newblock {\em Journal of Computational Physics}, 538:114155, 2025.

\bibitem{gallet2024two}
B.~Gallet.
\newblock Two-dimensional turbulence above topography: condensation transition and selection of minimum enstrophy solutions.
\newblock {\em Journal of Fluid Mechanics}, 988:A13, 2024.

\bibitem{holm1985nonlinear}
D.~D. Holm, J.~E. Marsden, T.~Ratiu, and A.~Weinstein.
\newblock Nonlinear stability of fluid and plasma equilibria.
\newblock {\em Physics reports}, 123(1-2):1--116, 1985.

\bibitem{hoppe1989diffeomorphism}
J.~Hoppe.
\newblock Diffeomorphism groups, quantization, and {$SU (\infty)$}.
\newblock {\em International Journal of Modern Physics A}, 4(19):5235--5248, 1989.

\bibitem{kahan1966accurate}
W.~Kahan.
\newblock Accurate eigenvalues of a symmetric tridiagonal matrix.
\newblock Technical Report CS41, Computer Science Department, Stanford University, Stanford, CA, July 1966.

\bibitem{kuo1959finite}
H.~Kuo.
\newblock Finite-amplitude three-dimensional harmonic waves on the spherical earth.
\newblock {\em Journal of Atmospheric Sciences}, 16(5):524--534, 1959.

\bibitem{lacasce2024vortices}
J.~H. Lacasce, A.~Pal{\'o}czy, and M.~Trodahl.
\newblock Vortices over bathymetry.
\newblock {\em Journal of Fluid Mechanics}, 979:A32, 2024.

\bibitem{luesink2024geometric}
E.~Luesink, A.~Franken, S.~Ephrati, and B.~Geurts.
\newblock Geometric derivation and structure-preserving simulation of quasi-geostrophy on the sphere.
\newblock {\em arXiv preprint arXiv:2402.13707}, 2024.

\bibitem{modin2025spatio}
K.~Modin and M.~Roop.
\newblock Spatio-temporal {Lie--Poisson} discretization for incompressible magnetohydrodynamics on the sphere.
\newblock {\em IMA Journal of Numerical Analysis}, page draf024, 2025.

\bibitem{modin2020casimir}
K.~Modin and M.~Viviani.
\newblock A {C}asimir preserving scheme for long-time simulation of spherical ideal hydrodynamics.
\newblock {\em Journal of Fluid Mechanics}, 884:A22, 2020.

\bibitem{modin2026brief}
K.~Modin and M.~Viviani.
\newblock A brief introduction to matrix hydrodynamics.
\newblock {\em Journal of Computational Dynamics}, 14(0):17--35, 2026.

\bibitem{morrison1986paradigm}
P.~J. Morrison.
\newblock A paradigm for joined {H}amiltonian and dissipative systems.
\newblock {\em Physica D: Nonlinear Phenomena}, 18(1-3):410--419, 1986.

\bibitem{priya2026two}
V.~K. Priya, S.~S. Patil, K.~Seshasayanan, and R.~Lakkaraju.
\newblock Two-dimensional turbulence over topography of varying roughness.
\newblock {\em Journal of Fluid Mechanics}, 1033:A33, 2026.

\bibitem{rhines1975waves}
P.~B. Rhines.
\newblock Waves and turbulence on a beta-plane.
\newblock {\em Journal of Fluid Mechanics}, 69(3):417--443, 1975.

\bibitem{roop2025thermal}
M.~Roop and S.~Ephrati.
\newblock Thermal quasi-geostrophic model on the sphere: Derivation and structure-preserving simulation.
\newblock {\em Physics of Fluids}, 37(9), 2025.

\bibitem{salmon1982geostrophic}
R.~Salmon.
\newblock Geostrophic turbulence.
\newblock {\em Topics in ocean physics}, 30:78, 1982.

\bibitem{salmon1976equilibrium}
R.~Salmon, G.~Holloway, and M.~C. Hendershott.
\newblock The equilibrium statistical mechanics of simple quasi-geostrophic models.
\newblock {\em Journal of Fluid Mechanics}, 75(4):691--703, 1976.

\bibitem{sanson2010evolution}
L.~Z. Sans{\'o}n, A.~Gonz{\'a}lez-Villanueva, and L.~Flores.
\newblock Evolution and decay of a rotating flow over random topography.
\newblock {\em Journal of fluid mechanics}, 642:159--180, 2010.

\bibitem{schubert2009shallow}
W.~H. Schubert, R.~K. Taft, and L.~G. Silvers.
\newblock Shallow water quasi-geostrophic theory on the sphere.
\newblock {\em Journal of Advances in Modeling Earth Systems}, 1(2), 2009.

\bibitem{siegelman2023two}
L.~Siegelman and W.~R. Young.
\newblock Two-dimensional turbulence above topography: {V}ortices and potential vorticity homogenization.
\newblock {\em Proceedings of the National Academy of Sciences}, 120(44):e2308018120, 2023.

\bibitem{theiss2004equatorward}
J.~Theiss.
\newblock Equatorward energy cascade, critical latitude, and the predominance of cyclonic vortices in geostrophic turbulence.
\newblock {\em Journal of physical oceanography}, 34(7):1663--1678, 2004.

\bibitem{verkley2009balanced}
W.~T. Verkley.
\newblock A balanced approximation of the one-layer shallow-water equations on a sphere.
\newblock {\em Journal of the atmospheric sciences}, 66(6):1735--1748, 2009.

\bibitem{weisstein3j}
E.~W. Weisstein.
\newblock Wigner 3j-symbol.
\newblock From \textit{MathWorld} -- A Wolfram Web Resource. Accessed 2026-03-02.

\bibitem{zeitlin1991finite}
V.~Zeitlin.
\newblock Finite-mode analogs of 2{D} ideal hydrodynamics: {C}oadjoint orbits and local canonical structure.
\newblock {\em Physica D: Nonlinear Phenomena}, 49(3):353--362, 1991.

\bibitem{zeitlin2004self}
V.~Zeitlin.
\newblock Self-consistent finite-mode approximations for the hydrodynamics of an incompressible fluid on nonrotating and rotating spheres.
\newblock {\em Physical review letters}, 93(26):264501, 2004.

\bibitem{zeitlin2018geophysical}
V.~Zeitlin.
\newblock {\em Geophysical fluid dynamics: understanding (almost) everything with rotating shallow water models}.
\newblock Oxford University Press, 2018.

\end{thebibliography}

\appendix
\section{Gaunt coefficients}\label{app:gaunt_coefficients}
Below, we briefly elaborate on the triple product of spherical harmonics appearing in Section \ref{subsec:asymptotic_regimes}.

We let $Y_{l,m}(\varphi, \theta)$ denote the complex spherical harmonic of degree $l$ and order $m$, and will omit the coordinates henceforth.
The spherical harmonics are eigenfunctions of the Laplace--Beltrami operator and satisfy $\Delta Y_{l,m} = -l(l+1)Y_{l,m}$.
The standard $L^2$ inner product on  complex valued functions on the sphere is denoted by angled brackets $\langle\cdot, \cdot\rangle$. That is, for any two functions $f, g$ on the sphere we have \begin{equation}
    \langle f, g\rangle = \int_{0}^{2\pi}\!\int_{0}^\pi f(\varphi, \theta)g^*(\varphi, \theta)\,\sin\theta\,\td\theta\,\td\varphi,
\end{equation}
where $g^*$ denotes the complex conjugate of $g$.
The spherical harmonics form an orthonormal basis for functions on the sphere with the provided inner product. 
The expansion in this basis will be used repeatedly, and given a general function $f$ on the sphere we adopt the following notation for its spectral coefficients, \begin{equation}
    f_{l,m} := \langle f, Y_{l,m}\rangle.
\end{equation}

Gaunt coefficients are the inner product of three spherical harmonics. These can be concisely expressed in terms of 3$j$-symbols, \begin{equation}
    \begin{split}
        \langle Y_{l_1,m_1} Y_{l_2,m_2}, Y_{l_3,m_3}\rangle & = \int\! Y_{l_1,m_1} Y_{l_2,m_2} Y_{l_3,m_3}\,\sin\theta\,\td\theta\td\phi \\
        &= \sqrt{\frac{(2 l_1 + 1)(2 l_2 + 1)(2 l_3 + 1)}{4\pi}}\begin{pmatrix}
            l_1 & l_2 & l_3 \\ 0 & 0 & 0
        \end{pmatrix}
        \begin{pmatrix}
            l_1 & l_2 & l_3 \\ m_1 & m_2 & m_3
        \end{pmatrix}.
    \end{split}
    \label{eq:gaunt}
\end{equation}
Here, the expressions in round brackets are the $3j$-symbols.
Below, we expand on their appearance throughout the paper by using the properties of 3$j$-symbols \cite{weisstein3j}.

The Gaunt coefficients appearing in the linear system \eqref{eq:linear_system_steady_state} satisfy $l_1 \in \{1, 2\}$ and $m_1=0$, which will lead to the following simplifications:
\begin{itemize}
    \item Since $m_1=0$, Eq. \eqref{eq:gaunt} is nonzero only when $m_3 = -m_2$;
    \item Due to the first $3j$-symbol, with all lower entries equal to zero, $l_1 + l_2 + l_3$ must be an even integer for the Gaunt coefficient to be nonzero. 
    \item In general, \eqref{eq:gaunt} is nonzero only when $|l_2 - l_1|\leq l_3\leq l_2 + l_1$ is satisfied. 
    Since $l_1\in \{1, 2\}$, this means that $\psi_{l', m'}$ will be coupled to coefficients of $\psi$ with $l$-values between $l'-2$ and $l'+2$, and topography coefficients between $l'-1$ and $l'+1$.
    This indicates that the linear system will be banded and sparse.
\end{itemize}
Using these simplifications, we find \begin{equation}
    \begin{split}
        \langle Y_{2, 0}\psi, Y_{l, m}\rangle &= \sum_{l', m'}\psi_{l', m'} \langle Y_{2,0} Y_{l', m'}, Y_{l, m}\rangle \\
        &= \sum_{l'}\psi_{l', -m}\langle Y_{2,0} Y_{l', -m}, Y_{l, m}\rangle \\
        &= \sum_{\substack{l'=l-2 \\ l'+l \text{~even}}}^{l+2} \psi_{l', -m}\langle Y_{2,0} Y_{l', -m}, Y_{l, m}\rangle,
    \end{split}
\end{equation}
which suggests that each $\psi_{l, m}$ will be directly coupled to at most three other spectral coefficients. In addition, $\langle Y_{2, 0} \psi, Y_{l, m}\rangle$ will only produce a term containing $\psi_{l, m}$ when $m=0$.
Analogously, we have \begin{equation}
    \langle Y_{1,0} h, Y_{l, m}\rangle = \sum_{\substack{l'=l-1 \\ l'+l\text{~odd}}}^{l+1}h_{l', -m}\langle Y_{1,0}Y_{l', -m}, Y_{l, m}\rangle.
\end{equation}

\section{Derivation of $F_C$}\label{app:nonlinear_stability_form}
Below, we provide the derivation of the positive-definite quadratic form $F_C$ defined in Equation \eqref{eq:pos_def_quad_form}.
We begin by writing out the definition of $F_C$,
\begin{equation*}
    \begin{split}
    F_C(\eta) &= 
    \mathcal{E}[\psi_* + \eta]-\mathcal{E}[\psi_*] + \lambda_*(E[\psi_* + \eta]-E_*) \\
    &= \frac{1}{2}\int\!\left(H[\psi_*+\eta] + \frac{2\mu}{\Ro}-\mu h\right)^2
     - \left(H[\psi_*] + \frac{2\mu}{\Ro}-\mu h\right)^2\\
     &
     \quad\quad - \lambda_* \Big((\psi_*+\eta)H[\psi_*+\eta] - \psi_*H[\psi_*]\Big)\,\td A.
    \end{split}
\end{equation*}
The expression above is simplified using the linearity of $H$, i.e., $H[\psi_*+\eta] = H[\psi_*] + H[\eta]$, and the relation $H[\psi_*] + 2\mu/\Ro - \mu h = q_*$, giving \begin{equation*}
    \begin{split}
      \frac{1}{2}\int\! \left(H[\eta] + q_*\right)^2 - q_*^2 - \lambda_* \left(\eta H[\psi_*] + \psi_* H[\eta] + \eta H[\eta] \right)\,\td A.
    \end{split}
\end{equation*}
The operator $H$ is self-adjoint, and we thus have $\int\!\eta H[\psi_*]\,\td A = \int\! \psi_* H[\eta]\,\td A$.
Upon expanding the square, we find \begin{equation*}
    \begin{split}
        &\frac{1}{2}\int\! \left( H[\eta]\right)^2 + 2 H[\eta]q_* + q_*^2 - q_*^2 - 2\underbrace{\lambda_*\psi_*}_{q_*} H[\eta] - \lambda_*\eta H[\eta]\,\td A \\ 
        & = \frac{1}{2}\int\! \left( H[\eta]\right)^2 - \lambda_* \eta H[\eta]\,\td A.
    \end{split}
\end{equation*}
The expression above is used in Equation \eqref{eq:pos_def_quad_form}.
We note that the expression above may be further expanded by using the definition $H[\eta]=\Delta \eta - \gamma\mu^2\eta$.
Via integration by parts of $\lambda_*\eta H[\eta]$, we obtain the expression \begin{equation*}
    F_C(\eta) = \frac{1}{2}\int\! \left( H[\eta]\right)^2 + \lambda_* |\nabla \eta|^2 + \lambda_* \gamma\mu^2\eta^2\,\td A.
\end{equation*}

\section{Validation of matrix size}\label{app:matrix_size}
The matrix size in Zeitlin's method can be considered the numerical resolution of the discretization.
We provide a validation of the adopted matrix size $N$ throughout the reported numerical tests in Section \ref{sec:numerical_demonstrations} by performing a grid refinement study.
In particular, we investigate how the choice of $N$ influences the Lagrange multiplier $\lambda$.

\begin{figure}[ht]
    \centering
    \includegraphics[width=0.5\linewidth]{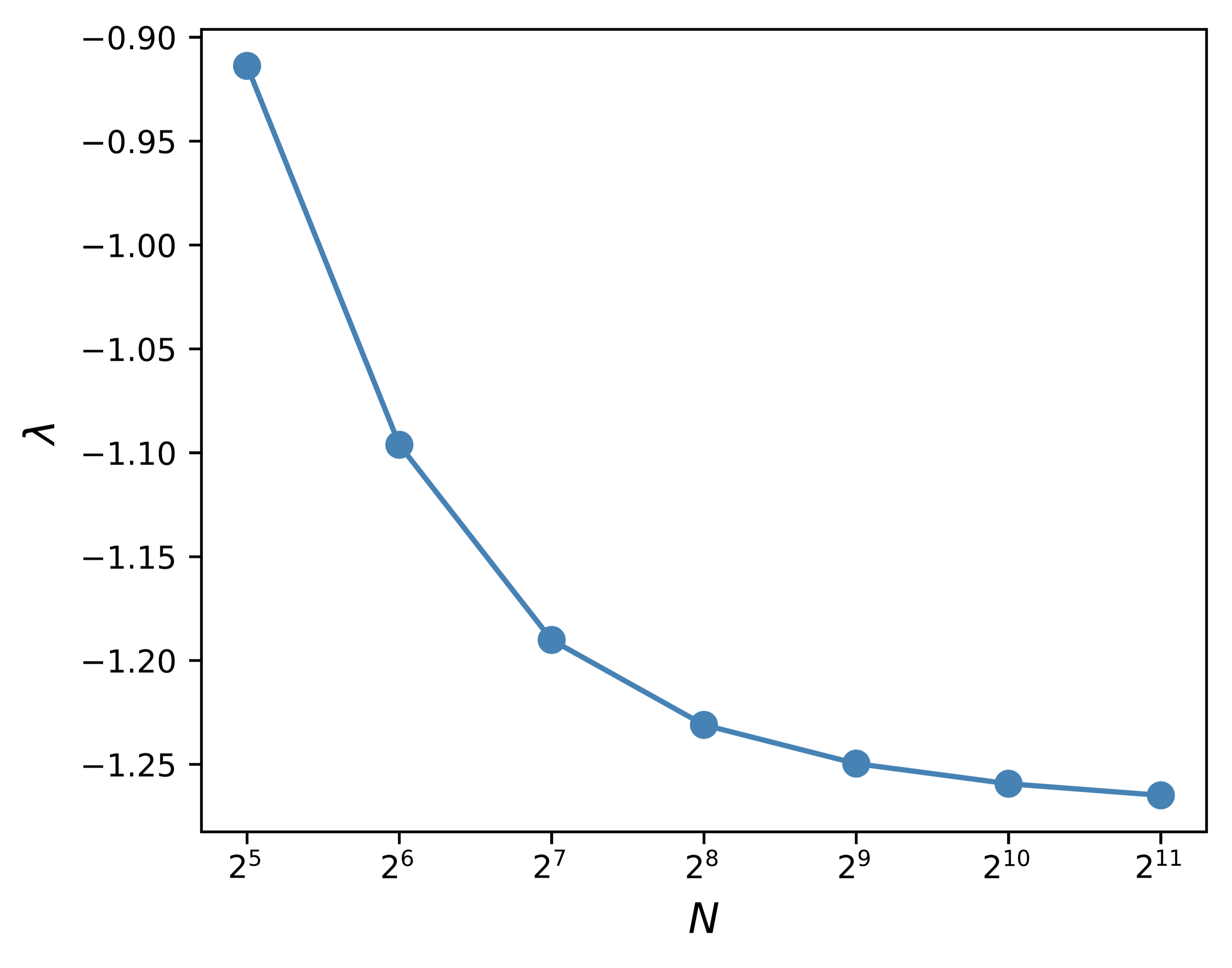}
    \caption{Log-linear plot of minimum-enstrophy Lagrange multiplier $\lambda$ against matrix size $N$.}
    \label{fig:lambda}
\end{figure}

We fix $\Ro = 10$ and $\gamma = 5$.
As the definition of energy \eqref{eq:zeitlin_energy} depends on $N$, we generate a random matrix $W \in \mathfrak{su}(2048)$ ($2048\times 2048$ skew-Hermitian matrices with zero trace) by 
\begin{align}
    \label{eq:matmat}
    W = \sum_{l=2}^{2048} \sum_{m=-l}^l a_{l,m} \frac{1}{(l^2+l+m+1)^\alpha} T_{l,m}^N
\end{align}
where $\alpha = 1.2$ and $a_{l,m}$  are independent, identically distributed standard normal random variables.
A random topography for $N=2048$ is generated in the same manner as the topography depicted in \Cref{fig:topography}.  
Then, for $N = 32, 64, 128, 256, 512, 1024$ and $2048$, $W$ is down-sampled by truncating the series \eqref{eq:matmat} at the appropriate number of terms and the energy is computed at the corresponding matrix size. 
We then solve the minimum-enstrophy problem at each $N$ to obtain $\lambda$.  
The found values of $\lambda$ are plotted against the matrix sizes in \Cref{fig:lambda}. 
We observe convergence of $\lambda$ as $N$ increases, and note that the relative change between $N= 1024$ and $N = 2048$ is small.
To reduce memory requirements, we adopt $N=1024$ in the numerical tests in Section \ref{sec:numerical_demonstrations}.

\end{document}